\documentclass[aip,reprint,nofootinbib,floatfix]{revtex4-2}

\usepackage{graphicx}

\usepackage{amsmath,amssymb,amsfonts}
\usepackage{dcolumn}
\usepackage{bm}
\usepackage{xcolor}
\usepackage{subcaption}
\usepackage{booktabs}
\usepackage{algorithm}
\usepackage{algorithmicx}
\usepackage{algpseudocode}
\usepackage{chemformula}
\usepackage{hyperref}
\usepackage{nicefrac}
\usepackage{microtype}

\newcommand{\chat}{\hat{c}}
\newcommand{\phihat}{\hat{\phi}}
\newcommand{\xhat}{\hat{x}}
\newcommand{\that}{\hat{t}}
\newcommand{\Lhat}{\hat{L}}
\newcommand{\pd}[2]{\frac{\partial #1}{\partial #2}}

\begin{document}

	\title{Physics-Informed Neural Networks for the Point Defect Model: Solving and Inverting Passive-Film Growth Kinetics}
	
	\author{Mohid Farooqi}
	\affiliation{Department of Physics and Astronomy, University of Waterloo,
		200 University Avenue West, Waterloo, Ontario N2L~3G1, Canada}
	
	\author{Ingmar B\"{o}sing}
	\affiliation{Chemical Process Engineering (CVT), University of Bremen,
		Bibliothekstra\ss{}e~1, 28359 Bremen, Germany}
	
	\author{Conrard Giresse Tetsassi Feugmo}
	\email{cgtetsas@uwaterloo.ca}
	\affiliation{Department of Chemistry, University of Waterloo, Waterloo, Ontario N2L 3G1, Canada}
	\affiliation{Department of Physics and Astronomy, University of Waterloo,
		200 University Avenue West, Waterloo, Ontario N2L~3G1, Canada}
	
	\date{\today}
	
	\begin{abstract}
		Predicting how passive oxide films grow and break down is
		central to corrosion science and the long-term integrity of structural
		alloys, and the Point Defect Model (PDM) is the standard kinetic
		description: a stiff, coupled system of Nernst--Planck, Poisson, and
		Butler--Volmer equations on a moving boundary. Solving it conventionally
		requires specialised finite-element (FEM) solvers, and identifying its
		kinetic parameters requires costly experimental campaigns. Physics-informed
		neural networks (PINNs) are compelling here because they natively
		assimilate data and invert for unknown parameters, so one measurement can
		replace an FEM solution and yield kinetics FEM cannot. We show
		that PINNs solve the PDM and recover its parameters from sparse data. The
		problem is difficult for reasons common to stiff multiphysics systems:
		widely separated scales, stiff boundary conditions, and convergence to
		non-physical solution branches. We address these with physics-based
		non-dimensionalisation, which extends stable simulation from about one hour
		to 250 hours; NTK adaptive weighting, which compresses a four-to-six-order
		loss imbalance to roughly one; and a single validated anchor that selects
		the physical branch and brings film-thickness error
		below $2.2\%$ at all five potentials. Stiff boundary-condition
		enforcement remains an open problem. Robustness comes from resampling the
		anchor each step, accuracy saturates beyond ten anchors, and the
		physics loss tolerates $5\%$ measurement noise. Crucially,
		recoverability tracks stiffness: a boundary-stiff kinetic constant is
		identifiable from film-thickness data while a weakly-coupled interior
		coefficient is not. By turning sparse measurements into full fields and
		inferred kinetics, the approach reduces the experimental and computational
		burden of characterising passive-film growth.
	\end{abstract}

	
	\maketitle
	
	\section{\label{sec:intro}Introduction}
	
	Passive oxide films (spontaneously formed nanometre-thin layers on metal
	surfaces) are the primary electrochemical barrier separating a metal from its
	environment~\cite{singh_harsh_2016,fu2021catalytic-646,ma2019origin-e88,
		maurice2018current-96f}.  Their growth kinetics determine the long-term
	corrosion protection of structural alloys in demanding applications including
	biomedical implants and nuclear reactor pressure vessels~\cite{macdonald_history_2011,
		Iannuzzi2022,ALAMIERY2024102672}.  Predicting how these films nucleate, thicken,
	and stabilise under applied potential and changing electrolyte chemistry is a
	central problem in corrosion science, particularly for systems where
	experimental access to the film/metal interface is limited.
	
	The Point Defect Model (PDM), originally proposed by Macdonald and
	co-workers~\cite{macdonald_history_2011}, remains the most widely used
	theoretical framework for passive film growth kinetics.  In the PDM, film growth
	and dissolution are governed by the migration and interaction of charged point
	defects (cation vacancies, anion vacancies, and metal interstitials) driven by
	electric fields and concentration gradients at both the metal/film and
	film/solution interfaces.  Refined versions of the PDM~\cite{Seyeux.2013,
		Leistner.2013,bosing_modeling_2023,alexiadis2025modeling-ef0,Li_PDM,
		Kolotinskii_2023} replace the original assumption of a constant electric field
	with an explicit solution of Poisson's equation, making the system a fully
	coupled set of partial differential equations (PDEs): the Nernst--Planck
	equation for each defect species, Poisson's equation for the electric potential,
	Butler--Volmer kinetics at both interfaces, and an ordinary differential equation
	(ODE) for the moving film boundary.  Solving this stiff multiphysics system has
	required specialized finite element solvers~\cite{bosing_modeling_2023,
		Kolotinskii_2023,engelhardt2024estimation-16a,bataillon2012numerical-9fe,
		sun2022point-49d}, which are accurate but expensive for parameter sweeps and
	difficult to adapt to inverse problems.
	
	Physics-informed neural networks (PINNs), introduced by Raissi, Perdikaris, and
	Karniadakis~\cite{raissi_physics-informed_2019}, embed the governing equations
	directly into the training loss function, enabling mesh-free solution of PDEs
	without labeled training data.  Two features make them well suited to the PDM in
	particular, and address exactly where FEM falls short: the network that solves
	the equations can simultaneously assimilate a data loss, and the governing
	constants can be made trainable, so a single PINN both fits sparse measurements
	and recovers unknown kinetic parameters rather than only producing a forward
	solution.  PINNs have been applied successfully to simpler
	electrochemical problems: Chen and co-workers demonstrated their use for
	voltammetric responses~\cite{chen2022predicting-abc,chen2022critical-abc} and
	hydrodynamic voltammetry~\cite{chen_hydrodynamic_2022}.  PINNs have also been
	extended to coupled phase-field equations with moving
	boundaries~\cite{kathane_physics_2024} and to phase-field models using
	NTK-based adaptive weighting~\cite{chen2025pfpinns-abc}.  Tutorials on PINNs
	for phase-field models have recently appeared in APL Machine
	Learning~\cite{li_tutorials_2024}, and comprehensive reviews of PINN
	architectures and failure modes are available~\cite{toscano_pinns_2024,
		sophiya_comprehensive_2025}.
	
	Despite this progress, PINNs have not yet been demonstrated on the full PDM with
	its combination of moving boundaries, stiff Butler--Volmer kinetics, and
	multi-scale variable ranges.  Our systematic investigation reveals four
	specific failure modes that prevent purely physics-informed learning from
	reaching quantitative accuracy on this class of problems.  The first
	two, scale disparity and loss imbalance, can be resolved with existing
	techniques.  The third, stiff boundary enforcement, resists all strategies we
	tested.  The fourth, convergence to non-physical solution branches, is resolved
	by adding one supervised data point.  We also report two strategies that did
	not work (Augmented Lagrangian methods and residual network connections) and
	explain why, as this negative knowledge is as valuable to the community as
	reporting what succeeds.
	
	The methodological contributions (non-dimensionalization, NTK-based adaptive
	weighting, hybrid anchoring, and loss-landscape diagnostics) are designed to
	transfer to any stiff coupled transport--reaction system.  The PDM serves
	as the validation benchmark.
	
	\section{\label{sec:pdm}The Point Defect Model}
	
	\subsection{Physical model}
	
	We follow the Refined PDM (R-PDM) of B\"{o}sing~\cite{bosing_modeling_2023},
	which resolves the electric potential explicitly and includes both cation
	vacancies (CV, charge $z_\text{CV} = -8/3$) and anion vacancies (AV, charge
	$z_\text{AV} = +2$).  The fractional cation-vacancy charge arises from a
	mean-field treatment of mixed Fe$^{2+}$/Fe$^{3+}$ site occupancy in the iron
	oxide lattice: one vacancy compensates $\tfrac{2}{3}$ of an Fe$^{2+}$ site and
	$\tfrac{1}{3}$ of an Fe$^{3+}$ site, giving $z_\text{CV} = -(\tfrac{2}{3}{\cdot}2
	+ \tfrac{1}{3}{\cdot}3) = -8/3$~\cite{bosing_modeling_2023}.  For simplicity,
	electronic carriers are excluded from the PINN
	implementation~\cite{bosing_modeling_2023}.  The interfacial
	reactions for iron passivation in halide-free solution are illustrated in
	Fig.~\ref{fig:PDM_Fe_defects} and listed in full in the Supporting Information
	(SI,~Section~S1).  The stoichiometry of chemical dissolution (R5) gives
	$2\,\text{Fe}^{3+} + \text{Fe}^{2+} + 4\,\text{H}_2\text{O}$, consistent
	with B\"{o}sing~\cite{bosing_modeling_2023}.
	
	\subsection{Governing equations}
	
	The oxide film occupies the domain $[0,\,L(t)]$, where $L(t)$ is the film
	thickness at time $t$.  Defect transport for species $i\in\{\text{CV,\,AV}\}$
	is governed by the Nernst--Planck equation,
	\begin{equation}
		\pd{C_i}{t} = -\nabla\cdot\mathbf{J}_i,\qquad
		\mathbf{J}_i = -D_i\nabla C_i - \frac{z_i F D_i C_i}{RT}\nabla\phi,
		\label{eq:NP}
	\end{equation}
	where $C_i$ (mol\,m$^{-3}$) is the molar concentration of species $i$, $D_i$
	(m$^2$\,s$^{-1}$) its diffusion coefficient, $z_i$ its charge number, $F =
	96485$~C\,mol$^{-1}$ the Faraday constant, $R = 8.314$~J\,mol$^{-1}$\,K$^{-1}$
	the gas constant, $T$ the absolute temperature, and $\phi$ (V) the electric
	potential.  The potential satisfies Poisson's equation,
	\begin{equation}
		-\nabla\cdot(\epsilon\nabla\phi) = F\sum_i z_i C_i,
		\label{eq:Poisson}
	\end{equation}
	where $\epsilon$ (F\,m$^{-1}$) is the dielectric permittivity of the oxide.
	Interfacial reaction rates $k_j$ follow Butler--Volmer kinetics,
	\begin{equation}
		k_j = k_j^0\exp\!\left(-\alpha_j\frac{F}{RT}\eta_j\right),
		\label{eq:BV}
	\end{equation}
	where $k_j^0$ is the pre-exponential rate constant, $\alpha_j$ the charge
	transfer coefficient, and $\eta_j = E - E_j^0$ the overpotential for reaction
	$j$, with $E$ the applied potential (V) and $E_j^0$ the equilibrium potential
	(V).  The single-exponential form retains only the dominant direction for each
	interfacial reaction: under the anodic conditions of passive film growth, metal
	oxidation at the metal/film interface and dissolution at the film/solution
	interface are both strongly irreversible ($|\eta_j| \gg RT/F$), so the
	back-reaction term is negligible~\cite{bosing_modeling_2023}.  Film growth couples to the defect fluxes via
	\begin{equation}
		\frac{dL}{dt} = \Omega\sum_j \nu_j k_j,
		\label{eq:film_growth}
	\end{equation}
	where $\Omega$ (m$^3$\,mol$^{-1}$) is the molar volume of the oxide and $\nu_j$
	the stoichiometric volume contribution of reaction $j$.  All boundary and
	initial conditions, together with all parameter values, are given in the SI
	(Section~S3).
	
\begin{figure*}[t]
		\centering
		\includegraphics[width=\textwidth]{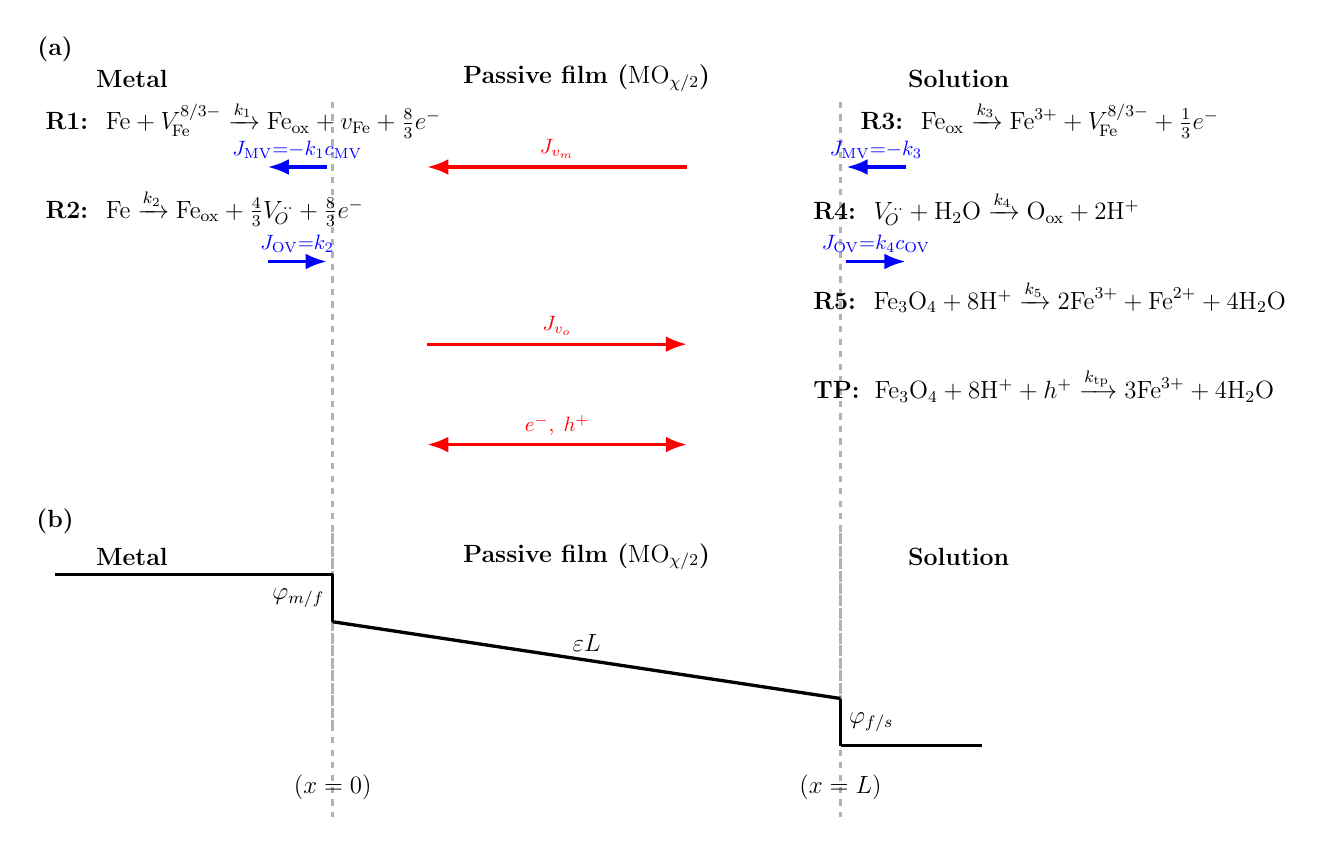}
		\caption{\label{fig:PDM_Fe_defects}%
			(a)~PDM defect reactions at the metal/film and film/solution interfaces
			for iron passivation in halide-free solution.
			(b)~Schematic potential distribution across the Fe/passive film/solution
			interface.  Species notation and boundary conditions are defined in the
			text.}
	\end{figure*}

	Figure~\ref{fig:PDM_Fe_defects}(a) illustrates the defect reactions at the
	metal/film (m/f) and film/solution (f/s) interfaces.  Species notation:
	Fe, iron atom in the metal; Fe$_\mathrm{ox}$, iron atom in the oxide lattice;
	$V_\mathrm{Fe}^{8/3-}$, iron cation vacancy; $V_O^{..}$, doubly-charged oxygen
	vacancy; $e^-$ and $h^+$, electron and hole carriers; Fe$^{n+}$, iron ion in
	solution.  The flux boundary conditions at the f/s interface are
	$J_\mathrm{CV}(x{=}L) = -k_3$ and $J_\mathrm{AV}(x{=}L) = k_4 c_\mathrm{AV}$.
	Panel~(b) shows the potential distribution across the Fe/passive film/solution
	interface, with drops $\varphi_{m/f}$ and $\varphi_{f/s}$ localised at the
	two boundaries.
	
	\section{\label{sec:method}PINN Methodology}
	
	\subsection{\label{subsec:nondim}Non-dimensionalization}
	
	Scale disparity is the first obstacle for PINNs applied to the PDM: spatial
	coordinates are on the order of one~nanometre ($10^{-9}$~m), concentrations
	span many orders of magnitude, and simulation times can reach $10^5$~s.  In
	32-bit floating-point arithmetic, gradients at nanometre length scales fall
	below machine precision, causing the network to learn flat profiles instead of
	the correct spatial variation (Fig.~\ref{fig:potential_demo}).
	
	We remove these disparities by rescaling all variables with characteristic
	scales grounded in the physics of the system:
	\begin{align}
		L_c &= 10^{-9}~\text{m}
		\quad(\text{one nanometre, initial film thickness}),
		\label{eq:Lc}\\
		t_c &= \frac{L_c^2}{D_\text{CV}}
		\quad(\text{diffusive time scale}),
		\label{eq:tc}\\
		c_c &= 10^{-5}~\text{mol\,m}^{-3}
		\quad(\text{reference concentration}),
		\label{eq:cc}\\
		\phi_c &= \frac{RT}{F} \approx 0.0257~\text{V at }298~\text{K}
		\quad(\text{thermal voltage}).
		\label{eq:phic}
	\end{align}
	The choice $\phi_c = RT/F$ is natural for electrochemical systems because it
	makes the Butler--Volmer exponents dimensionless with coefficients equal to the
	charge transfer coefficient times the charge number, both of order one.  The
	dimensionless variables are
	\begin{equation}
		\xhat = \frac{x}{L_c},\quad
		\that = \frac{t}{t_c},\quad
		\chat_i = \frac{C_i}{c_c},\quad
		\phihat = \frac{\phi}{\phi_c},\quad
		\Lhat = \frac{L}{L_c}.
		\label{eq:nondim_vars}
	\end{equation}
	With this scaling, all variables remain within $\mathcal{O}(1)$ to
	$\mathcal{O}(100)$ throughout the domain, which is the range where floating-point
	arithmetic is reliable.  The practical benefit is dramatic: dimensional models
	become numerically unstable at around 3600~s, whereas the non-dimensional model
	runs stably to $900\,000$~s, a factor of 250 improvement in temporal range.
	The non-dimensional governing equations, boundary conditions, and the
	derived dimensionless parameters (Péclet numbers, Damköhler number, and
	screening ratio) are given in full in the SI (Section~S2).
	
	\subsection{\label{subsec:arch}Segregated network architecture}
	
\begin{figure*}[t]
		\centering
		\includegraphics[width=0.82\textwidth]{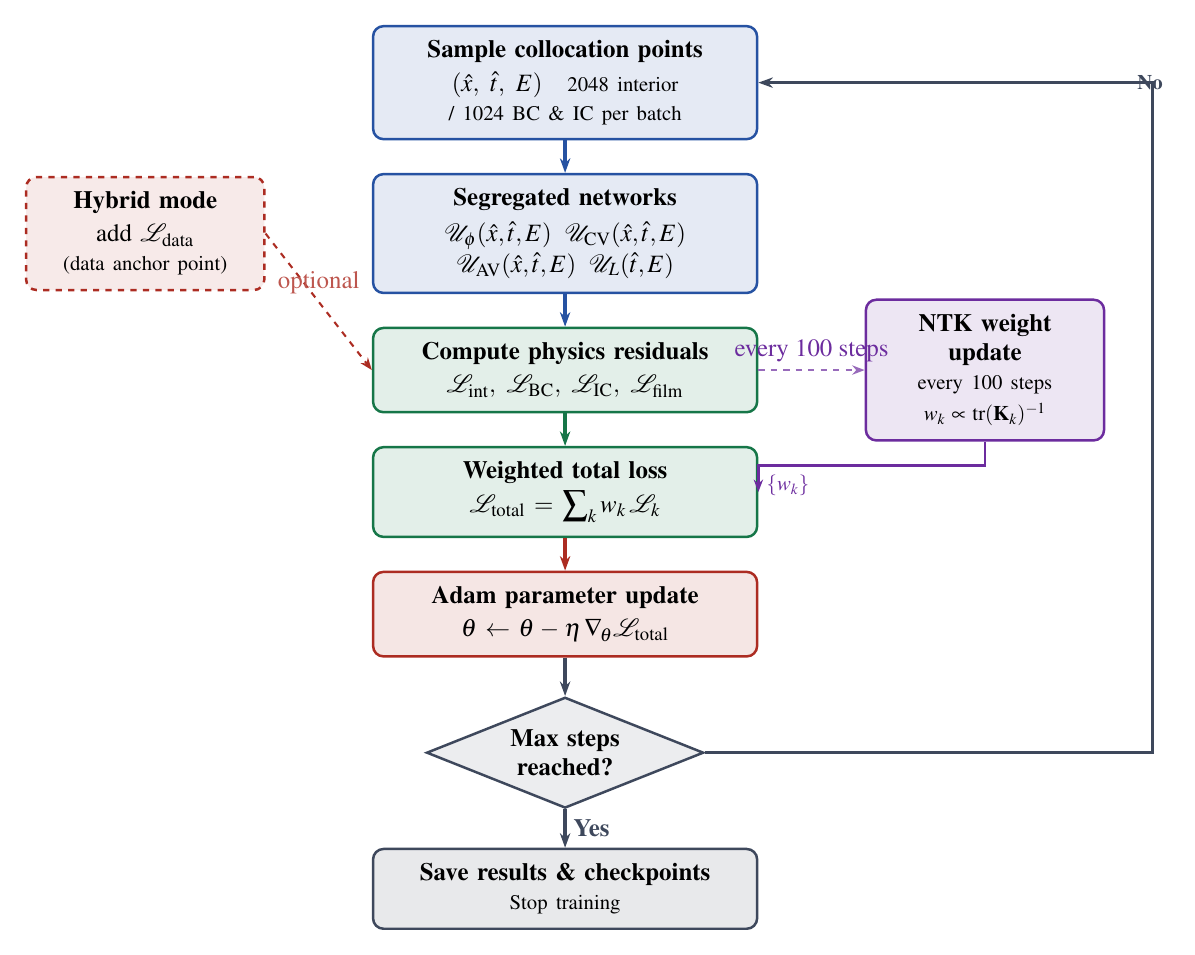}
		\caption{\label{fig:arch}%
			PINNACLE training loop with four segregated networks and NTK adaptive
			loss weighting.  The dashed red path indicates hybrid mode, in which a
			single FEM anchor point is added to the loss.}
	\end{figure*}

	Assigning a single network to represent all fields concentrates conflicting
	gradient signals in one set of weights.  Following the approach used for coupled
	phase-field~\cite{kathane_physics_2024} and microfluidic
	systems~\cite{sun_physics-informed_2024}, we use a \textit{segregated}
	architecture in which each physical field has its own dedicated network.
	Specifically, $\mathcal{U}_\phi(\xhat,\that,E)$ predicts the electric potential,
	$\mathcal{U}_\text{CV}(\xhat,\that,E)$ the cation vacancy concentration,
	$\mathcal{U}_\text{AV}(\xhat,\that,E)$ the anion vacancy concentration, and
	$\mathcal{U}_L(\that,E)$ the film thickness.  The spatial input $\xhat$ is
	omitted from $\mathcal{U}_L$ because film thickness depends only on time and
	the applied potential, not on position within the film.
	Each network uses five hidden layers of 20~neurons with Swish activation
	functions~\cite{al-safwan_is_2021}, chosen for its smooth gradient properties
	relative to tanh.  The four field networks together hold
	approximately $9\mathrm{k}$ trainable parameters; this $5\times20$ depth and
	width follows the configuration recommended by Chen et
	al.~\cite{chen2025pfpinns-abc} for NTK-weighted PINNs on coupled-field
	problems, and preliminary tests over the range $4\times20$ to $6\times32$
	produced no statistically significant change in the converged hybrid error.
	The applied potential $E$ is sampled once per mini-batch and
	shared across all collocation points in that batch, correctly encoding the
	experimental reality that $E$ is spatially uniform.  This also means the trained
	model can sweep the entire polarisation curve in a single inference pass without
	retraining.  The PINNACLE training loop is shown in Fig.~\ref{fig:arch}.
	At each step, collocation points $(\hat{x},\hat{t},E)$ are sampled and passed
	to the four segregated networks.  Their outputs are used to evaluate the PDE,
	BC, IC, and film-growth residuals.  Every 100 steps, NTK adaptive weights are
	recomputed and the loss components are rebalanced before the Adam parameter
	update is applied.  In hybrid mode, a single FEM-validated film-thickness point
	is included as an anchor loss term $\mathcal{L}_\mathrm{data}$, weighted
	identically to the film-growth loss.
	
	\subsection{\label{subsec:loss}Loss function}
	
	The training loss combines PDE residuals (interior), boundary conditions (BC),
	initial conditions (IC), and a film-growth ODE term.  For each state variable
	$j \in \{\text{CV, AV, }\phi\}$, the interior loss penalises the discrepancy
	between the predicted time derivative and the PDE operator $\mathcal{N}_j$:
	\begin{align}
		\mathcal{L}_\text{int} &= \sum_{j}\frac{1}{N_\text{int}}
		\sum_{i=1}^{N_\text{int}}
		\left|\mathcal{N}_j\!\bigl(\mathcal{U}_j(x_i,t_i)\bigr)
		- \pd{\mathcal{U}_j(x_i,t_i)}{t}\right|^2,
		\label{eq:lint}\\
		\mathcal{L}_\text{BC} &= \sum_{j}\frac{1}{N_\text{BC}}
		\sum_{i=1}^{N_\text{BC}}
		\left|\partial\Omega_j\!\bigl(\mathcal{U}_j(x_i,t_i)\bigr)\right|^2,
		\label{eq:lbc}\\
		\mathcal{L}_\text{IC} &= \sum_{j}\frac{1}{N_\text{IC}}
		\sum_{i=1}^{N_\text{IC}}
		\left|\mathcal{I}_j\!\bigl(\mathcal{U}_j(x_i)\bigr)\right|^2,
		\label{eq:lic}\\
		\mathcal{L}_\text{film} &= \frac{1}{N_\text{film}}
		\sum_{i=1}^{N_\text{film}}
		\left|\pd{\mathcal{U}_L(t_i)}{t} - \frac{dL}{dt}(t_i)\right|^2,
		\label{eq:lfilm}\\
		\mathcal{L}_\text{total} &= w_\text{int}\mathcal{L}_\text{int}
		+ w_\text{BC}\mathcal{L}_\text{BC}
		+ w_\text{IC}\mathcal{L}_\text{IC}
		+ w_\text{film}\mathcal{L}_\text{film}.
		\label{eq:ltotal}
	\end{align}
	The scalar weights $w_k$ are updated adaptively using the NTK framework
	described in Sec.~\ref{subsec:ntk}.
	
	\subsection{\label{subsec:ntk}NTK-based adaptive loss balancing}
	
	Even after non-dimensionalization, the Poisson loss dominates the transport
	losses by four to six orders of magnitude under uniform weighting
	(Fig.~\ref{fig:loss_panels}(a,b)).  When this happens, the optimizer effectively
	ignores the minority-loss terms and the PINN learns only the dominant equation.
	
	The Neural Tangent Kernel (NTK) framework provides a principled way to balance
	this.  Introduced by Jacot, Gabriel, and Hongler~\cite{jacot_neural_2020}, the
	NTK describes how a neural network's outputs change with respect to small
	parameter perturbations during training.  In the infinite-width limit, training
	is equivalent to kernel gradient descent with a fixed kernel.  For finite
	networks, the NTK provides a useful approximation: the eigenvalues of the NTK
	matrix for each loss term determine the effective learning rate of that term.
	If one term's NTK eigenvalues are much larger than another's, the optimizer
	updates that term aggressively and neglects the other, the
	mechanism by which Wang et al.~\cite{wang2022respecting} explain why PINNs
	often fail to train.  Equalising the NTK
	traces across loss components therefore equalizes the effective learning rates.
	
	Concretely, let $\bm{\theta} \in \mathbb{R}^M$ denote the network parameter
	vector.  For loss component $k$ evaluated at $N_k$ collocation points
	$\{(x_n, t_n, E_n)\}_{n=1}^{N_k}$, define the per-point residual Jacobian
	matrix
	\begin{equation}
		\mathbf{J}_k \in \mathbb{R}^{N_k \times M},\qquad
		(\mathbf{J}_k)_{nm}
		= \frac{\partial\, r_k(x_n,t_n,E_n)}{\partial\theta_m},
		\label{eq:Jac}
	\end{equation}
	where $r_k(x_n,t_n,E_n)$ is the per-point residual.  The NTK matrix is the
	outer product
	\begin{equation}
		\mathbf{K}_k = \mathbf{J}_k\mathbf{J}_k^\top \in \mathbb{R}^{N_k\times N_k},
		\label{eq:NTK}
	\end{equation}
	and its trace,
	\begin{equation}
		\mathrm{tr}(\mathbf{K}_k)
		= \sum_{n=1}^{N_k}\sum_{m=1}^{M}
		\left(\frac{\partial\,r_k(x_n,t_n,E_n)}{\partial\theta_m}\right)^2,
		\label{eq:trace}
	\end{equation}
	measures the total sensitivity of component $k$ to parameter changes.  We set
	\begin{equation}
		w_k(s) = \frac{N_k^{-1}\,\mathrm{tr}(\mathbf{K}_k(s))}
		{\displaystyle\sum_{j} N_j^{-1}\,\mathrm{tr}(\mathbf{K}_j(s))},
		\label{eq:NTKweights}
	\end{equation}
	following the formulation of Chen et al.~\cite{chen2025pfpinns-abc}.  Computing
	the full Jacobian at every step is prohibitive ($\mathcal{O}(N_k \times M)$
	memory per component), so we use a mini-batch approximation: by the
	central limit theorem~\cite{muller_achieving_2023}, the diagonal mean of
	$\mathbf{K}_k$ over a random sub-batch of size $b_k$ converges to the true mean
	as $b_k \to N_k$.  We select $b_k$ such that the coefficient of variation of
	the estimate is below 0.2~\cite{chen2025pfpinns-abc}, and recompute weights
	every 100 training steps for computational efficiency.

	The realised sub-batch sizes are $b_\mathrm{int}=1024$,
	$b_\mathrm{BC}=512$, $b_\mathrm{IC}=512$, and $b_\mathrm{film}=1024$.
	Recomputing the NTK weights every 100 steps raises the mean per-step training
	cost from $34.8$~ms under uniform weighting to $48.5$~ms, and the peak GPU
	memory from $237$ to $805$~MB: an overhead of roughly $14$~ms/step and
	$570$~MB that buys the loss balancing above. Two numerical-stability measures
	are also applied. First, per-component spike rejection: any step whose
	film-physics, boundary, interior, or initial loss component exceeds $10^{15}$
	is skipped, with its gradients zeroed and the Adam state left unchanged, which
	protects the optimiser from non-physical transients early in training without
	altering the kinetics. Second, the FEM data-matching term
	$\mathcal{L}_\mathrm{data}$ is added to the weighted physics sum without
	rescaling, so the anchor constraint is always enforced.
	
	\subsection{\label{subsec:hybrid}Hybrid training with minimal data}
	
	As discussed in Sec.~\ref{subsec:patho}, pure PINN
	training, despite converging successfully, settles onto the wrong solution
	branch of the PDM equations.  To resolve this, we add a single supervised data
	term
	\begin{equation}
		\mathcal{L}_\text{data} = \left|\mathcal{U}_L(t^*,E^*) - L^*\right|^2
		\label{eq:ldata}
	\end{equation}
	to the total loss, where $(t^*,\,E^*,\,L^*)$ is one FEM-validated point.  This
	term carries the same adaptive NTK weight as the film loss.
	The sensitivity to this anchor is quantified in
	Sec.~\ref{subsec:robust}: the result is reproducible when the anchor is
	resampled at each step, whereas a single fixed anchor is placement-sensitive.
	The PINN already encodes the correct physics and needs only the anchor to
	select the physical solution branch.
	Combining a sparse set of supervised observations with a
	physics loss is a well-studied strategy for inverse and data-scarce PDE
	problems: Raissi et al.~\cite{raissi2020hidden} demonstrated it for hidden
	fluid mechanics, Chen, Liu, and Sun~\cite{chen2021scarce} recovered governing
	equations from as few as $\mathcal{O}(10^2)$ observations, and recent work has
	quantified the data requirements for surrogate modelling of stirred
	tanks~\cite{travnikova2025quantifying} and 3D flow--thermal
	problems~\cite{bhatnagar2024pinn}; Cuomo et al.~\cite{cuomo2022scientific}
	give a broad survey.
	
	The complete training procedure is summarised in Algorithm~\ref{alg:pinnacle}.
	
	\begin{algorithm}[H]
		\caption{PINNACLE: PINN training with NTK adaptive loss balancing}
		\label{alg:pinnacle}
		\begin{algorithmic}[1]
			\State \textbf{Input:} networks $\{\mathcal{U}_\phi,\mathcal{U}_\text{CV},\mathcal{U}_\text{AV},\mathcal{U}_L\}$, parameter vector $\bm{\theta}$
			\State \textbf{Initialise:} Adam optimizer, $w_k\leftarrow 1$ for all $k$, step $s\leftarrow 0$
			\While{$s < s_\mathrm{max}$}
			\State Sample collocation points: $\mathcal{X}_\mathrm{int}$ (2048), $\mathcal{X}_\mathrm{BC}$ (1024), $\mathcal{X}_\mathrm{IC}$ (1024), $\mathcal{X}_\mathrm{film}$ (2048)
			\For{each point set $\mathcal{X}_k$}
			\State Compute predictions and PDE/BC/IC residuals $r_k$
			\State $\mathcal{L}_k \leftarrow \frac{1}{|\mathcal{X}_k|}\sum_i |r_k^{(i)}|^2$
			\EndFor
			\If{$s \bmod 100 = 0$}
			\For{each component $k$}
			\State Draw random sub-batch of size $b_k$ from $\mathcal{X}_k$
			\State Compute Jacobian $\mathbf{J}_k$ (per-point gradients w.r.t.\ $\bm{\theta}$) on sub-batch
			\State $\lambda_k \leftarrow \frac{1}{b_k}\,\mathrm{tr}(\mathbf{J}_k\mathbf{J}_k^\top)$
			\EndFor
			\State $w_k \leftarrow \lambda_k \big/ \sum_j \lambda_j$ \hfill\Comment{renormalize weights}
			\EndIf
			\State $\mathcal{L}_\mathrm{total} \leftarrow \sum_k w_k\mathcal{L}_k$
			\State $\bm{\theta} \leftarrow \bm{\theta} - \eta\,\nabla_{\bm{\theta}}\mathcal{L}_\mathrm{total}$ \hfill\Comment{Adam step}
			\State $s \leftarrow s+1$
			\EndWhile
			\State \textbf{Return:} trained networks $\{\mathcal{U}_\phi^*,\mathcal{U}_\text{CV}^*,\mathcal{U}_\text{AV}^*,\mathcal{U}_L^*\}$
		\end{algorithmic}
	\end{algorithm}
	
	\section{\label{sec:results}Results}
	
	\subsection{Benchmark setup}
	
	We validate against FEM solutions of the R-PDM computed by
	B\"{o}sing~\cite{bosing_modeling_2023} in COMSOL Multiphysics with adaptive
	meshing to handle the moving boundary.  The FEM boundary conditions and their
	convergence are documented in that reference.  We evaluate film thickness
	predictions across applied potentials $E \in [0.1,\,1.8]$~V and simulation
	times up to $9\times10^5$~s ($\approx$250~hours).  All PINN models are trained
	for 50\,000 Adam steps at a learning rate of $10^{-3}$.  For convergence
	verification, we additionally ran training at 100\,000 steps and confirmed that
	the hybrid error changes by less than 0.1~percentage point; the results reported
	here are converged.
	
	\subsection{Pure PINN results}
	
\begin{figure*}[t]
		\centering
		\includegraphics[width=\textwidth]{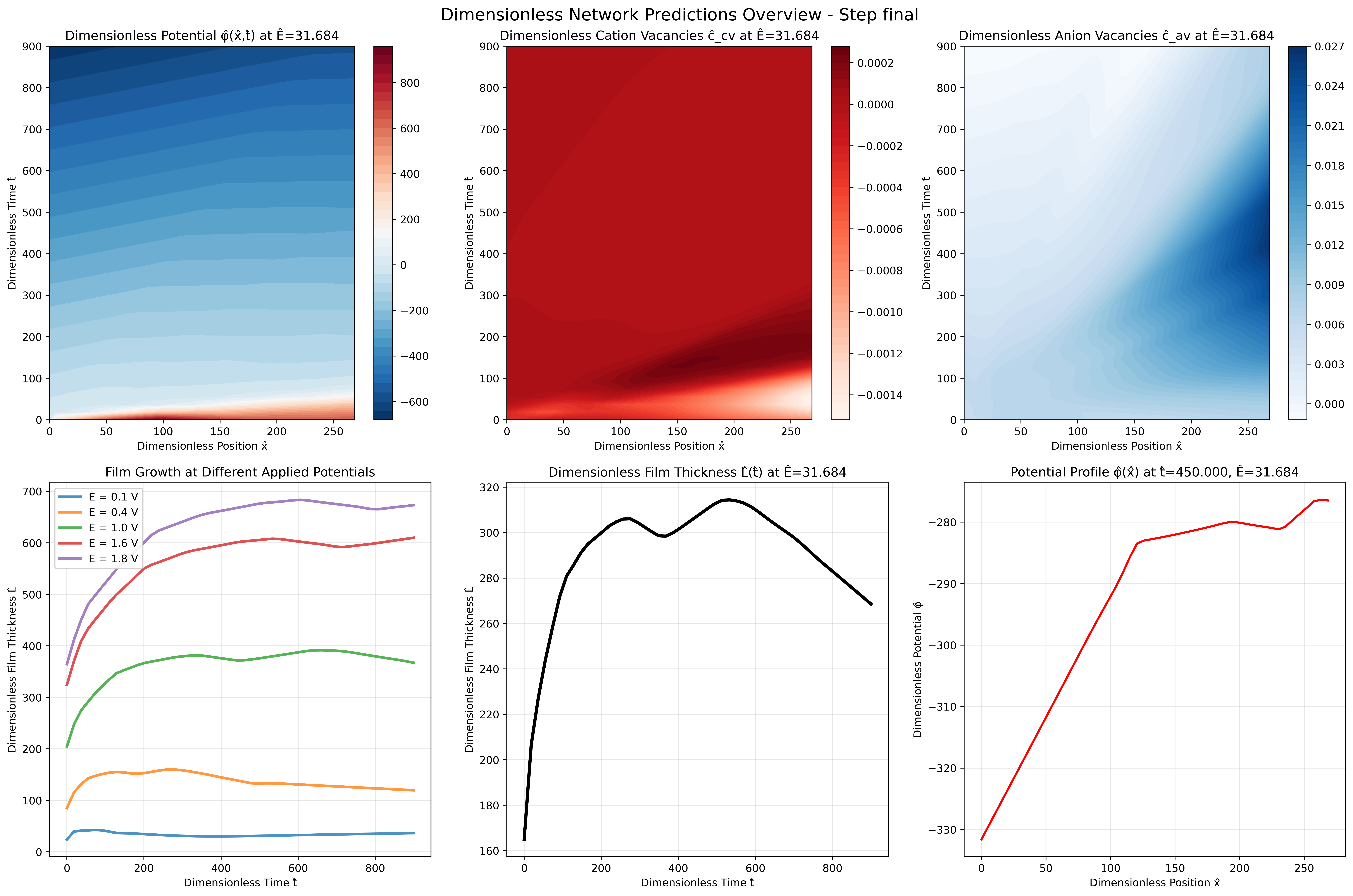}
		\caption{\label{fig:predictions_ntk}%
			Pure PINN predictions (NTK weighting, 250~hours).  Top row: dimensionless
			potential, cation vacancy, and anion vacancy profiles at selected times.
			Bottom row: film thickness versus time at representative potentials.
			Qualitative trends are physically plausible; absolute scale does not match
			the FEM (see Fig.~\ref{fig:pure_vs_fem}).}
	\end{figure*}

	Figure~\ref{fig:predictions_ntk} shows the field profiles predicted by the pure
	NTK-weighted PINN.  The model reproduces the correct qualitative behavior:
	film growth is exponential and voltage-dependent, and the potential and
	concentration profiles evolve consistently with the physical picture.  At
	$E = 0.1$~V, growth is minimal; at higher potentials the film grows faster, as
	expected from the Butler--Volmer kinetics.
	
\begin{figure*}[t]
		\centering
		\begin{subfigure}[t]{0.32\textwidth}
			\includegraphics[width=\textwidth]{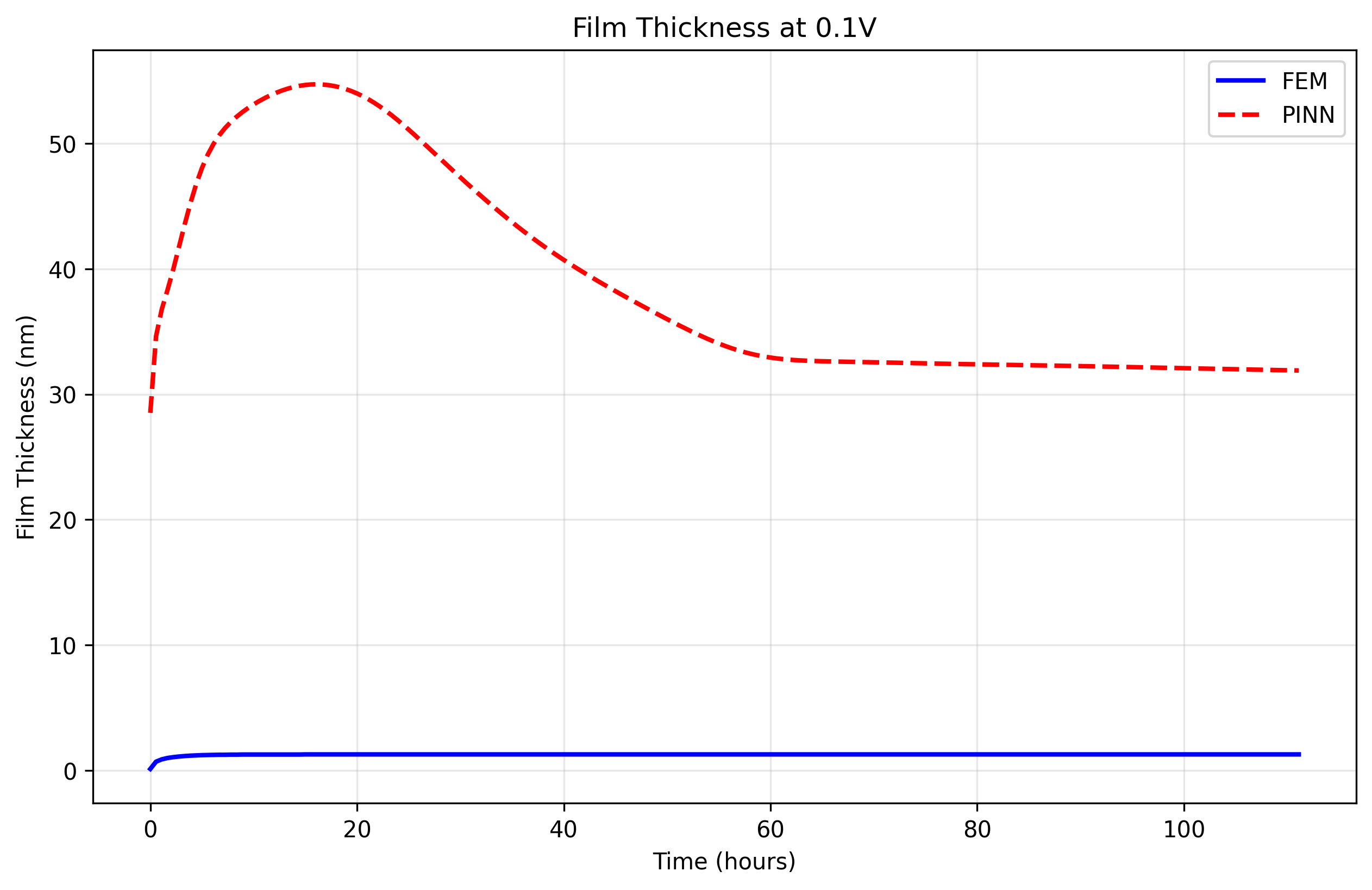}
			\caption{$E=0.1$~V}
		\end{subfigure}\hfill
		\begin{subfigure}[t]{0.32\textwidth}
			\includegraphics[width=\textwidth]{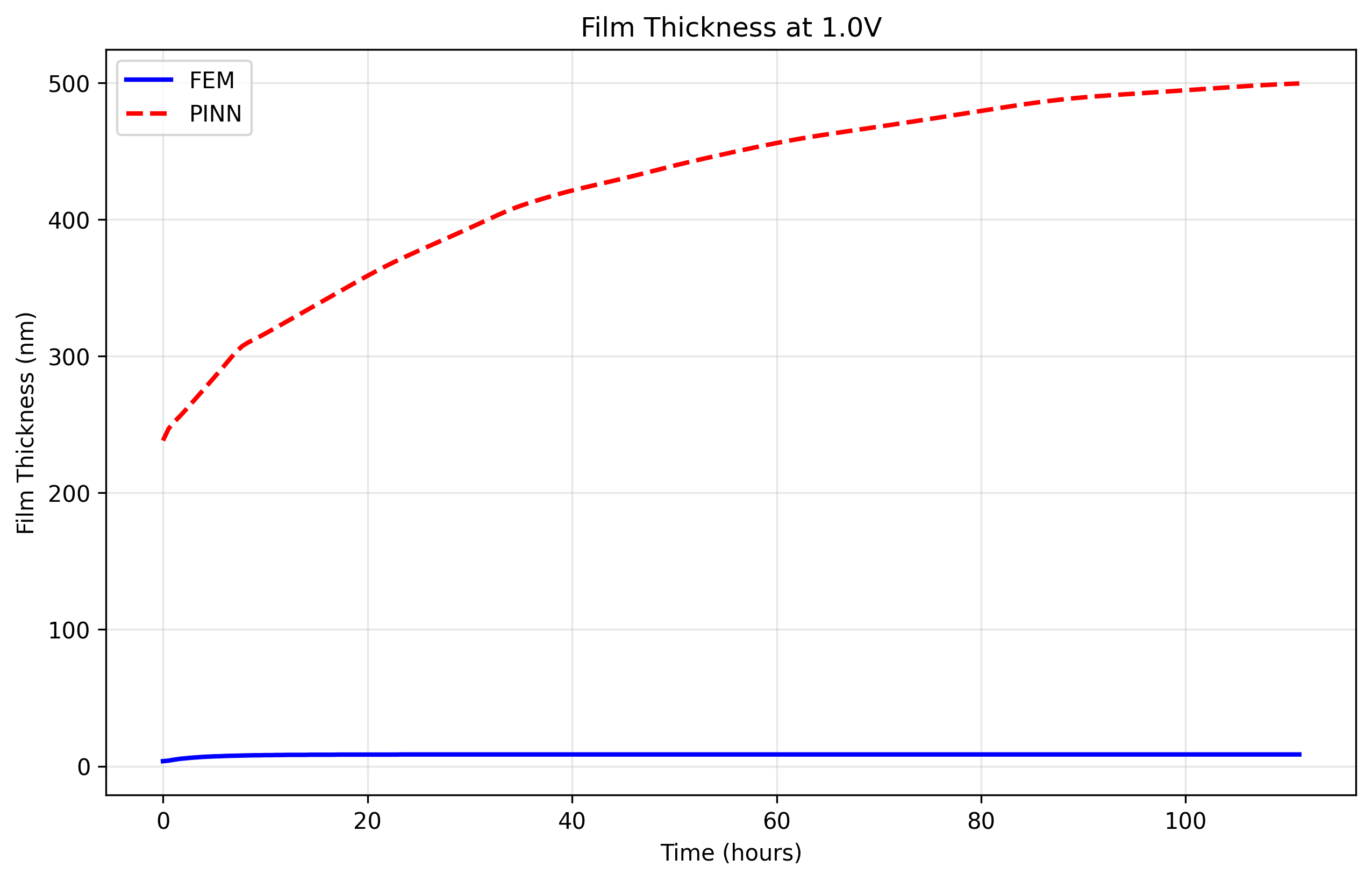}
			\caption{$E=1.0$~V}
		\end{subfigure}\hfill
		\begin{subfigure}[t]{0.32\textwidth}
			\includegraphics[width=\textwidth]{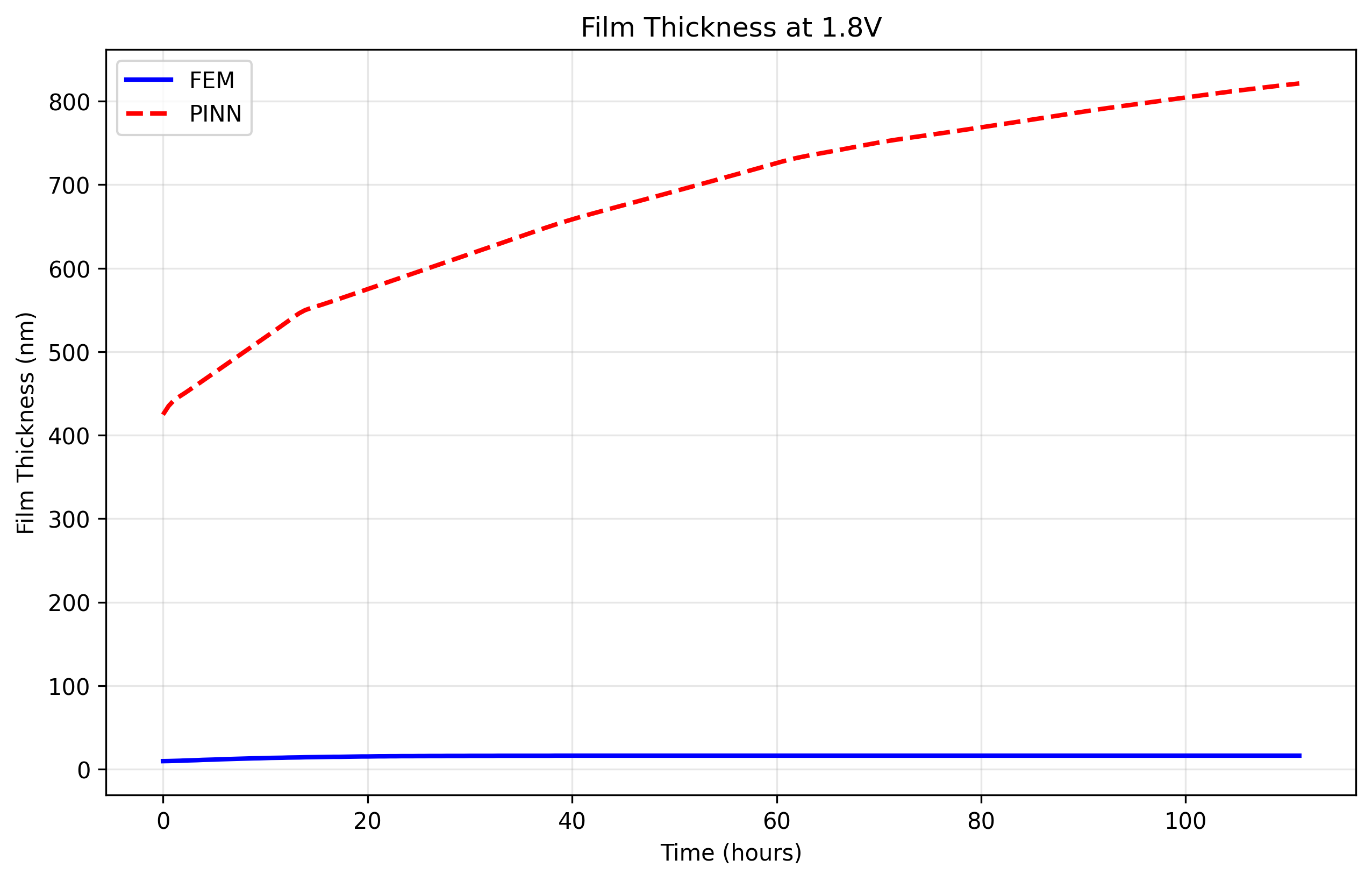}
			\caption{$E=1.8$~V}
		\end{subfigure}
		\caption{\label{fig:pure_vs_fem}%
			Pure PINN vs.\ FEM film thickness at three applied potentials.  The PINN
			captures the correct exponential shape and voltage ordering, but overestimates
			the absolute scale by factors of 20--50.}
	\end{figure*}

	However, a direct comparison with the FEM reference reveals a large quantitative
	discrepancy (Fig.~\ref{fig:pure_vs_fem}).  The PINN overestimates film
	thickness by one to two orders of magnitude at all voltages tested
	(Table~\ref{tab:comparison}).  At $E = 0.1$~V it predicts $32$~nm against the
	FEM reference of $1.27$~nm; at $E = 1.8$~V it predicts $821$~nm against
	$16.1$~nm.  This is not a training failure: losses converge cleanly and the
	predicted potential profiles are dimensionally plausible~\cite{bosing_modeling_2023}.
	Instead, we interpret this as convergence to a different branch of the solution
	manifold, as discussed in Sec.~\ref{sec:failure}C.
	
	\subsection{Hybrid PINN results}
	
\begin{figure*}[t]
		\centering
		\includegraphics[width=\textwidth]{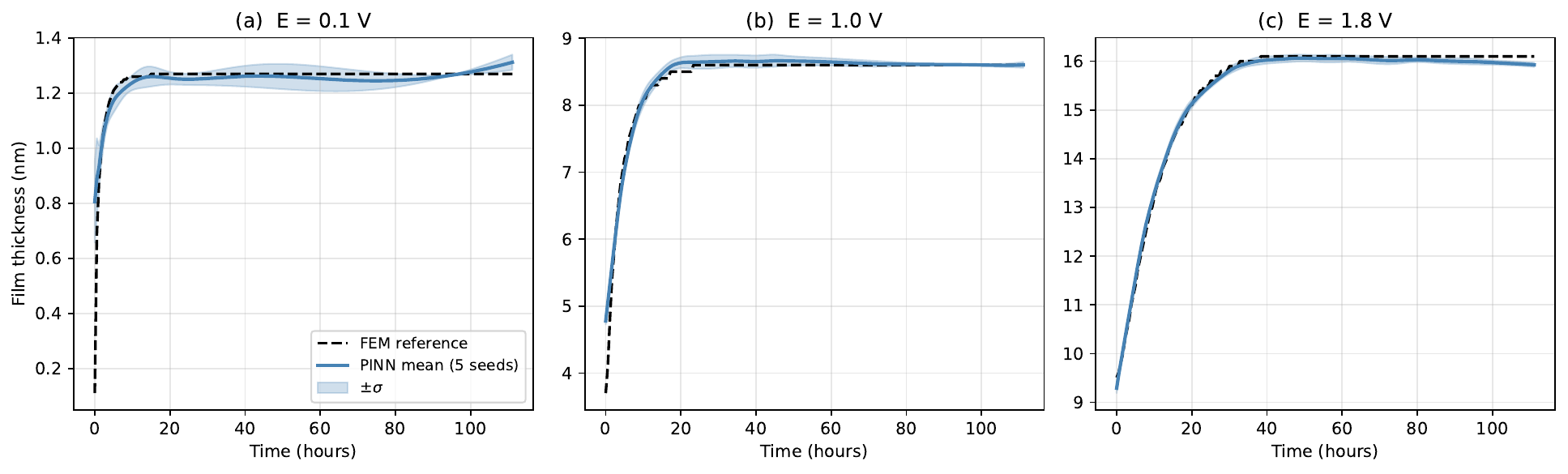}
		\caption{\label{fig:hybrid_robustness}%
			Hybrid PINN vs.\ FEM: a single anchor steers the PINN
			onto the physical solution branch at all voltages, and the result is
			reproducible. Mean PINN film-thickness prediction (solid) with
			$\pm\sigma$ band (shaded) over the five converged random-anchor seeds,
			against the FEM reference (dashed), at (a)~$0.1$~V, (b)~$1.0$~V, and
			(c)~$1.8$~V. The mean tracks FEM across the full transient, with the
			spread widest at the hardest low-voltage case.}
	\end{figure*}

	Adding a single FEM data point $(t^* = 150\,000$~s, $E^* = 0.1$~V,
	$L^* = 1.27$~nm$)$, chosen at random without optimisation, transforms the
	results dramatically.  The hybrid predictions match the FEM reference closely
	at every voltage tested (Fig.~\ref{fig:hybrid_robustness}).  At $E = 0.1$~V the
	error drops from $2412\%$ to $0.32\%$ and $R^2 = 0.90$; comparable improvements
	hold across the full voltage range (Table~\ref{tab:comparison}).
	
	The $R^2$ values are computed over the full transient time series, not just the
	final film thickness.  Lower values at intermediate voltages (e.g., $R^2 = 0.87$
	at $0.4$~V) reflect the fact that the FEM transient at those conditions is short:
	the initial mismatch during the first few thousand seconds carries proportionally
	more weight in the $R^2$ metric, while the final film thickness error at the same
	voltage is below $1\%$.
	
	\begin{table}[htbp]
		\caption{\label{tab:comparison}%
			Film-thickness error (final value) and $R^2$ over the full transient for the
			pure PINN and the hybrid PINN at five applied potentials.  The improvement
			factor is the ratio of pure to hybrid error.}
		\begin{ruledtabular}
			\begin{tabular}{crrrc}
				$E$ (V) & Pure (\%) & Hybrid (\%) & Improvement & $R^2$ \\
				\hline
				0.1 & 2412 & 0.32 & $7537\times$ & 0.8975 \\
				0.4 & 4707 & 0.97 & $4852\times$ & 0.8708 \\
				1.0 & 5708 & 0.80 & $7135\times$ & 0.9123 \\
				1.6 & 5218 & 2.18 & $2393\times$ & 0.9428 \\
				1.8 & 5001 & 0.38 & $13160\times$ & 0.9891 \\
			\end{tabular}
		\end{ruledtabular}
	\end{table}
	
	\subsection{Robustness to anchor placement}\label{subsec:robust}

	The hybrid result is robust in two senses that we quantify
	separately. First, it is reproducible across independent random-anchor
	streams: we repeated the hybrid training for seven independent random seeds,
	each resampling one FEM anchor point per step from the full solution. Six of
	the seven converge to the physical branch, with whole-curve relative
	film-thickness errors (averaged over the five test voltages) between $1.7\%$
	and $14.6\%$ (mean $4.3\%$, with five of the six clustering tightly at
	$2.3\pm0.5\%$; Fig.~\ref{fig:hybrid_robustness}); the remaining seed converges to a non-physical branch
	($240\%$), an initialisation-driven failure rather than a property of the
	anchor. An independent re-run of four of these seeds under a fixed
	$50{,}000$-step protocol reproduces the same statistics (three of four
	converge, $4.3\pm3.1\%$). Second, this robustness stems specifically from \emph{resampling} the
	anchor at every step, not from any single point being sufficient. When one
	FEM point is instead held fixed for the entire run, the error is large and
	strongly dependent on where that point is placed: a sweep over five fixed
	positions in the $(t,E)$ domain gives errors from $84\%$ to $353\%$
	(Table~\ref{tab:anchor_position}). The redraw-every-step scheme therefore
	behaves less like a single supervised label and more like a stochastic
	regulariser that repeatedly nudges the solution toward the physical branch.

	\begin{table}[htbp]
		\caption{\label{tab:anchor_position}%
			Single fixed-anchor position sweep. A lone FEM anchor is
			held fixed throughout training at the indicated $(t^*,E^*)$; the error is
			the whole-curve relative $L_2$ film-thickness error averaged over the
			five test voltages. Contrast with the $\sim$2--4\% obtained when the
			anchor is resampled every step.}
		\begin{ruledtabular}
			\begin{tabular}{lccr}
				Anchor & $t^*$ (ks) & $E^*$ (V) & Error (\%) \\
				\hline
				early, low $E$ & 2   & 0.1 & 353 \\
				mid, low $E$   & 200 & 0.1 & 84  \\
				late, low $E$  & 380 & 0.1 & 85  \\
				mid, mid $E$   & 200 & 1.0 & 104 \\
				mid, high $E$  & 200 & 1.6 & 341 \\
			\end{tabular}
		\end{ruledtabular}
	\end{table}

	\subsection{Data efficiency}

\begin{figure}[!t]
		\centering
		\includegraphics[width=\columnwidth]{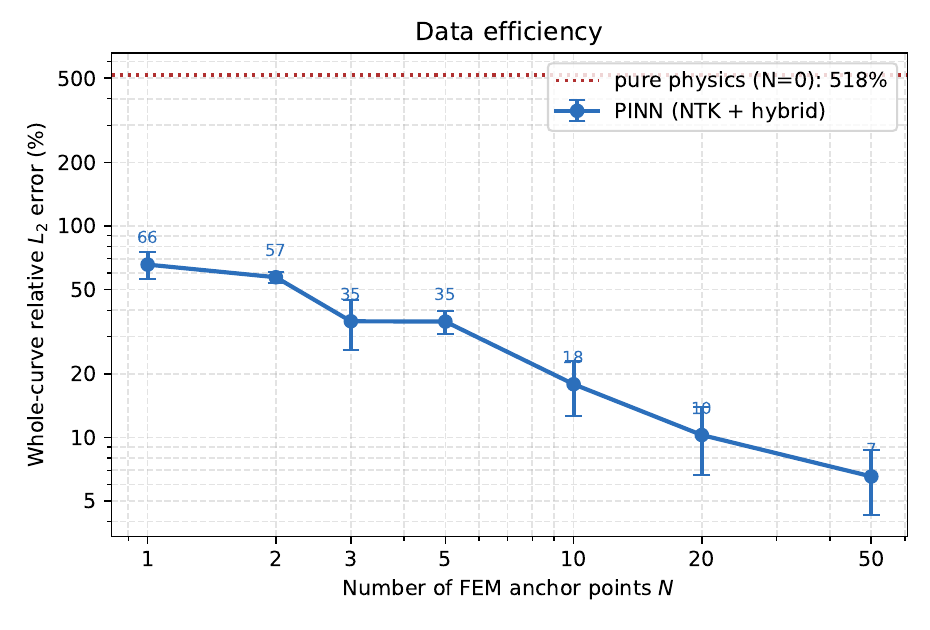}
		\caption{\label{fig:data_efficiency}%
			Data efficiency. Whole-curve relative $L_2$
			film-thickness error (mean $\pm$ s.d.\ over three seeds, averaged over
			the five test voltages) versus the number of fixed FEM anchor points
			$N_\mathrm{data}$, on log--log axes. Returns diminish beyond
			$N_\mathrm{data}\approx10$.}
	\end{figure}

	We next varied the number of fixed FEM anchor points
	$N_\mathrm{data}$ supplied during training, from $N_\mathrm{data}=0$ (pure
	PINN) up to $50$, averaging over three random seeds at each level
	(Fig.~\ref{fig:data_efficiency}). With no anchor the error is enormous and
	highly variable (mean $518\%$), confirming the branch ambiguity. A single
	fixed anchor already reduces the error to about $66\%$, and the error then
	falls steadily: $36\%$ at $N_\mathrm{data}=3$, $18\%$ at $N_\mathrm{data}=10$,
	and $6.5\%$ at $N_\mathrm{data}=50$. Returns diminish beyond roughly ten
	anchor points, so a handful of FEM evaluations already captures most of the
	available accuracy.

	\subsection{Noise robustness}

\begin{figure}[!t]
		\centering
		\includegraphics[width=\columnwidth]{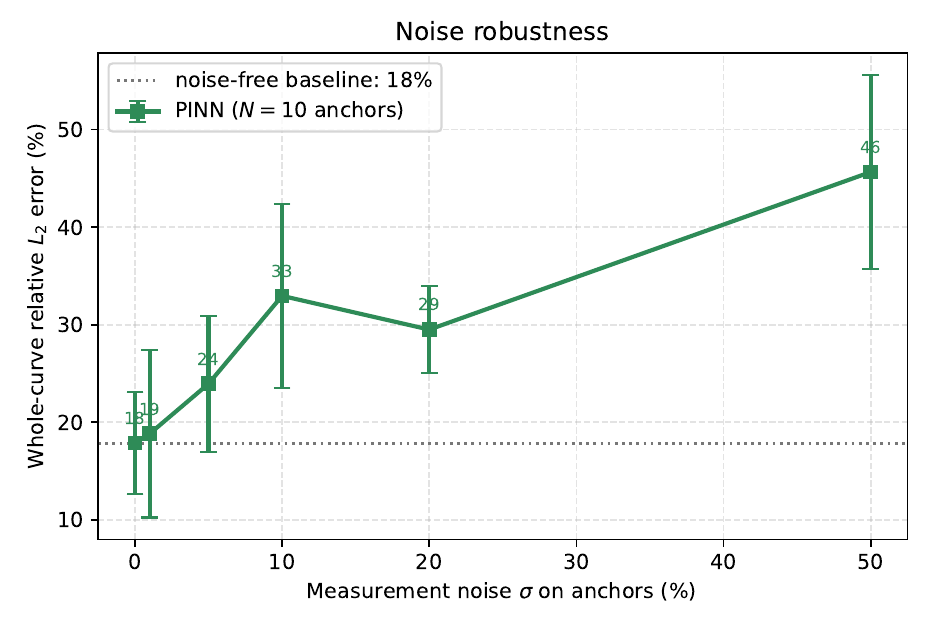}
		\caption{\label{fig:noise}%
			Noise robustness. Whole-curve relative $L_2$
			film-thickness error (mean $\pm$ s.d.\ over three seeds, averaged over
			the five test voltages) versus the standard deviation $\sigma$ of
			multiplicative Gaussian noise on the anchor measurement, at
			$N_\mathrm{data}=10$.}
	\end{figure}

	Experimental film-thickness data carry measurement
	uncertainty, so we tested how multiplicative Gaussian noise on the anchor
	value, $L^*\!\rightarrow L^*(1+\varepsilon)$ with
	$\varepsilon\sim\mathcal{N}(0,\sigma^2)$, propagates into the solution
	(Fig.~\ref{fig:noise}, $N_\mathrm{data}=10$, three seeds per level). The
	physics loss acts as a denoiser: the error rises only gradually, from
	$17.9\%$ at $\sigma=0$ to $23.9\%$ at $\sigma=5\%$, and remains usable
	($\sim$30--46\%) even at $\sigma=50\%$. The film-growth physics constrains the
	solution enough that a noisy anchor still selects the correct branch.

	\subsection{Inverse problem: parameter recovery}

	The same framework supports inverse problems. Rather than
	taking the kinetic parameters as known, we treat a chosen parameter as an
	unknown, represent it as a learnable variable in $\log_{10}$ space, and
	optimise it by gradient descent alongside the network weights against a small
	set of $L(t)$ observations. To keep the coupled solve stable we use a
	two-stage schedule: the networks are trained first with the parameter held at
	its initial guess, after which the networks are frozen and only the parameter
	is updated. We recover one parameter at a time, since the joint recovery of
	several kinetic constants from a single observable is ill-posed.

\begin{figure*}[t]
		\centering
		\includegraphics[width=0.9\textwidth]{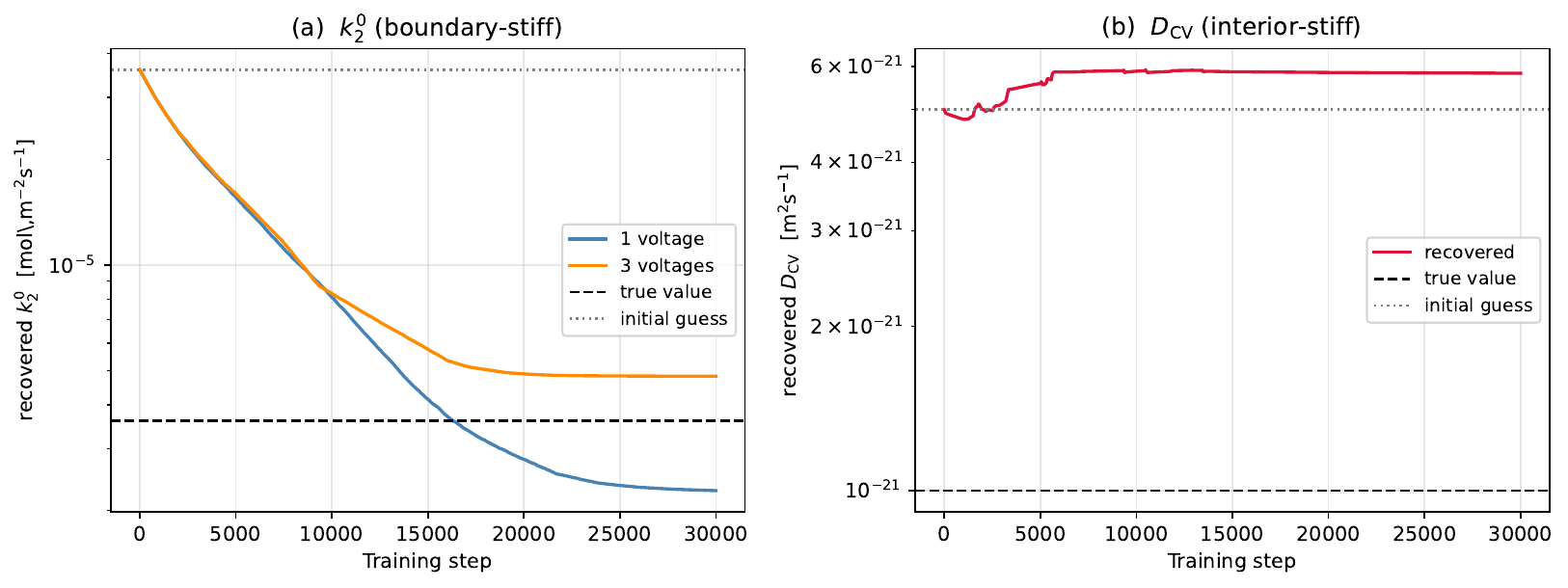}
		\caption{\label{fig:inverse}%
			Parameter recovery during training. (a)~The
			boundary-stiff rate constant $k_2^0$ descends monotonically from a
			ten-fold initial overestimate toward its true value (dashed) under
			observations at one and at three voltages, settling within about a third
			of the true value. (b)~The interior-stiff diffusion coefficient
			$D_\mathrm{CV}$ barely moves and drifts away from its true value
			(dashed, an order of magnitude below the initial guess), confirming that
			film-thickness data alone do not constrain it.}
	\end{figure*}

	The outcome is governed by which loss term the parameter is
	stiff in. The rate constant $k_2^0$ is \emph{boundary-stiff}: it enters the
	film-growth law, Eq.~\eqref{eq:film_growth}, directly through the exponential
	Butler--Volmer term, so it controls the observable $L(t)$ strongly. Starting
	from an initial guess ten times the true value
	($3.6\times10^{-5}$ against a true $3.6\times10^{-6}$), the optimiser drives
	$k_2^0$ monotonically down through more than a decade of error
	(Fig.~\ref{fig:inverse}) and settles
	within about a third of the true value: $2.3\times10^{-6}$ ($37\%$ error)
	from observations at a single voltage, and $4.8\times10^{-6}$ ($34\%$ error)
	when three voltages are observed. The residual error is set by the forward
	model's own $L(t)$ fit rather than by the number of observation voltages, and
	in both cases $k_2^0$ is recovered to the correct order of magnitude.

	The diffusion coefficient $D_\mathrm{CV}$ behaves oppositely.
	It is \emph{interior-stiff}: it appears in the Nernst--Planck transport
	equation, Eq.~\eqref{eq:NP}, and influences $L(t)$ only indirectly, through
	the vacancy flux that feeds the moving boundary. Starting from a guess five
	times the true value ($5\times10^{-21}$ against a true $1\times10^{-21}$), the
	parameter barely moves and in fact drifts further from the true value
	($5.8\times10^{-21}$, a $483\%$ error): the film-thickness observations simply
	do not constrain it. Recovering $D_\mathrm{CV}$ would require observations of
	the interior concentration field, not of $L(t)$ alone.

	These two cases show that parameter recoverability in this
	system tracks stiffness: a parameter that drives the observable directly is
	identifiable from sparse film-thickness data, whereas one that couples to the
	observable only weakly is not. This provides a practical diagnostic for any
	inverse PINN problem, indicating which parameters a given measurement can
	constrain before any optimisation is attempted.

	\section{\label{sec:failure}Failure Modes and Solutions}
	
	\subsection{Scale disparity and non-dimensionalization}
	
\begin{figure}[!t]
		\begin{subfigure}{0.48\columnwidth}
			\includegraphics[width=\textwidth]{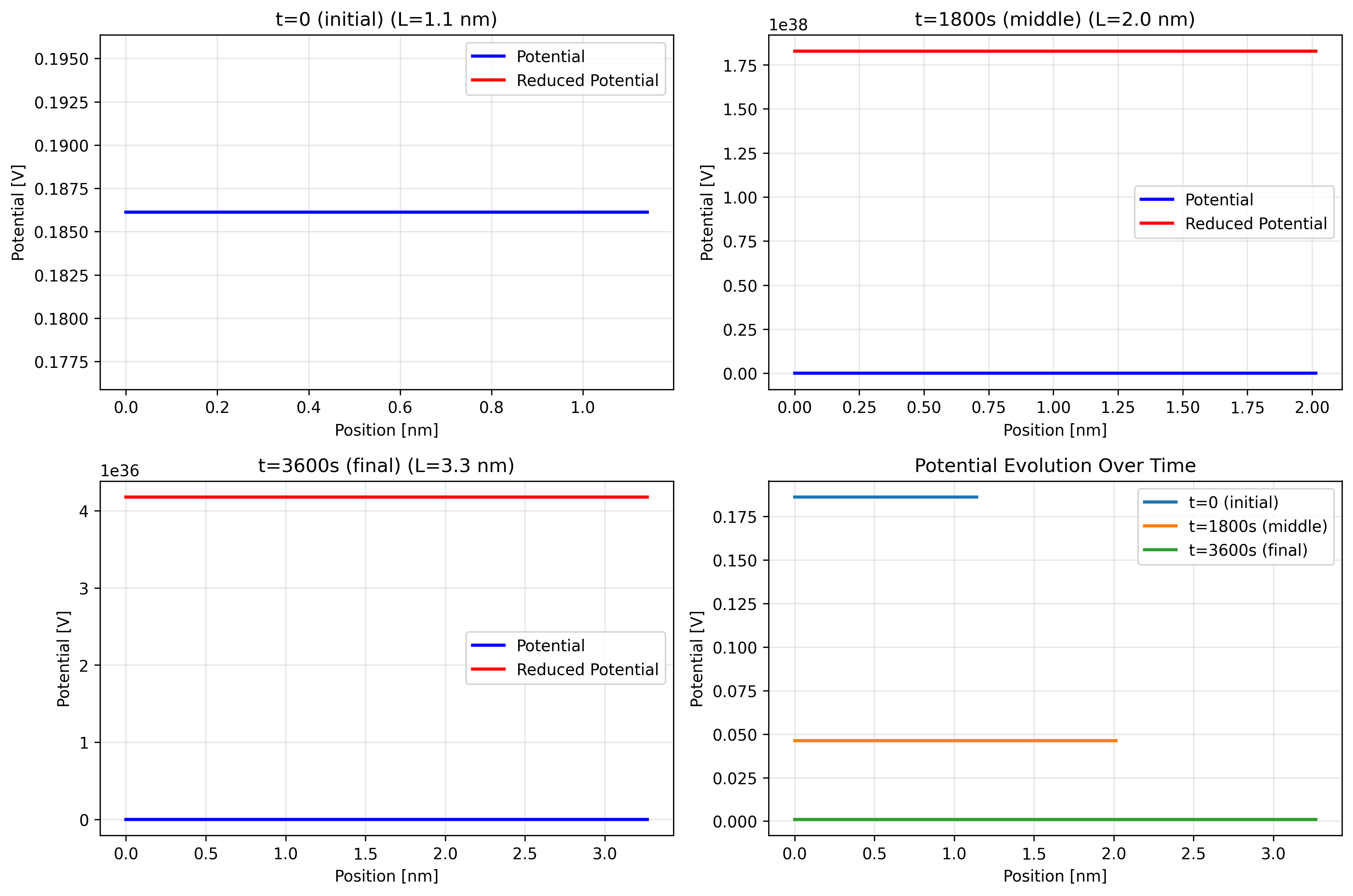}
			\caption{Dimensional model}
		\end{subfigure}\hfill
		\begin{subfigure}{0.48\columnwidth}
			\includegraphics[width=\textwidth]{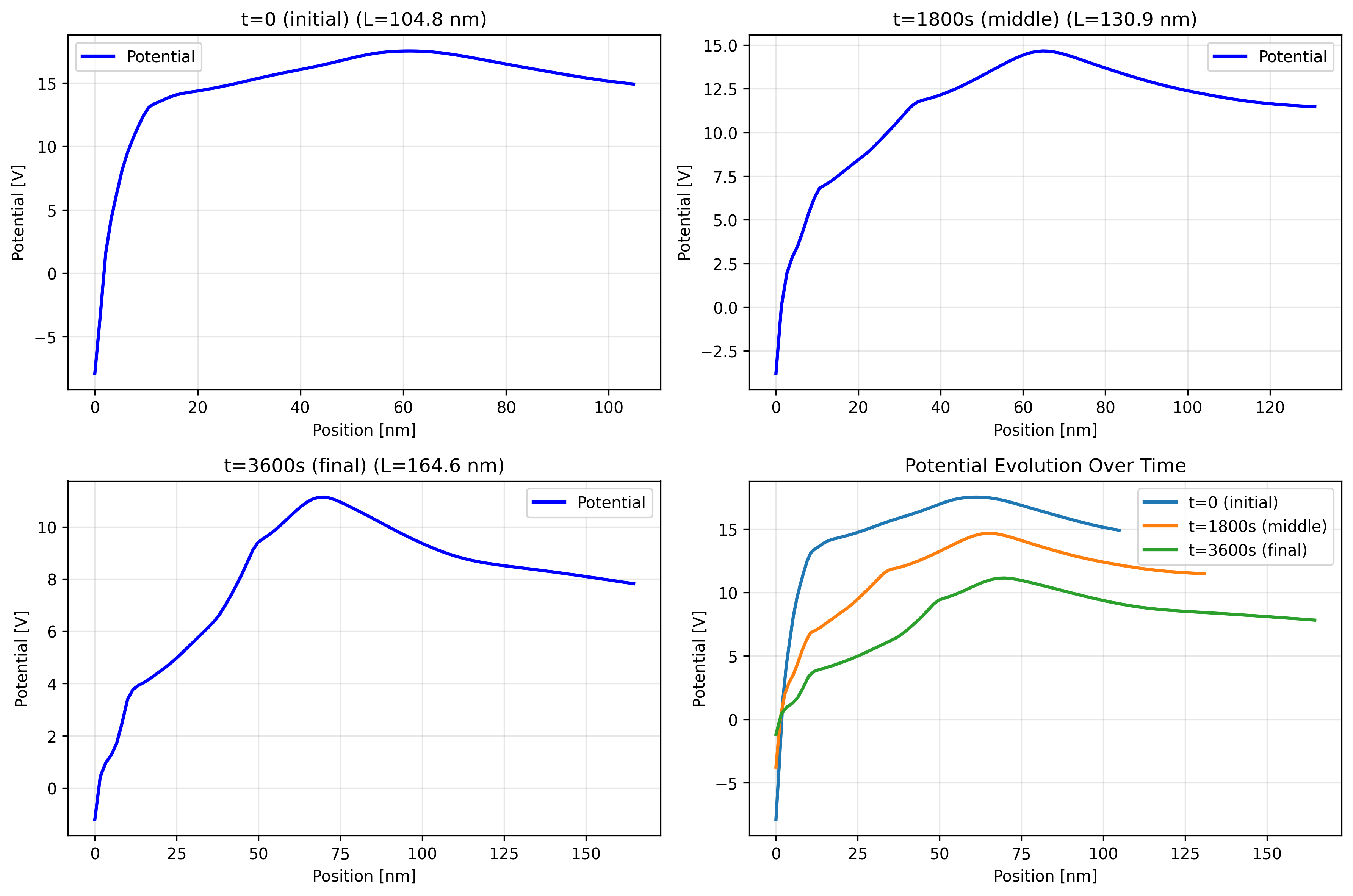}
			\caption{Non-dimensional model}
		\end{subfigure}
		\caption{\label{fig:potential_demo}%
			Potential profiles at representative times.  The dimensional model learns
			flat (constant) profiles due to floating-point underflow at nanometre scales.
			The non-dimensional model captures the correct spatial gradient.}
	\end{figure}

	Without non-dimensionalization, spatial coordinates ($\sim 10^{-9}$~m) and
	concentration differences fall below the resolution of 32-bit floating-point
	arithmetic.  The network then learns constant or nearly constant potential
	profiles (Fig.~\ref{fig:potential_demo}), in clear violation of the physical
	boundary conditions.  Furthermore, dimensional models become numerically
	unstable beyond about 3600~s, whereas the non-dimensional model remains stable
	to $9\times10^5$~s, a factor-of-250 improvement in temporal reach.
	
	Non-dimensionalization with physics-based characteristic scales
	(Eqs.~\eqref{eq:Lc}--\eqref{eq:phic}) resolves both problems.
	Malekjani et al.~\cite{malekjani_comparative_2025} confirm, for a similar
	multi-scale problem, that physics-based rescaling consistently outperforms
	statistical normalization.
	
	\subsection{Loss imbalance and NTK weighting}
	
\begin{figure*}[!t]
		\begin{subfigure}{0.32\textwidth}
			\includegraphics[width=\textwidth]{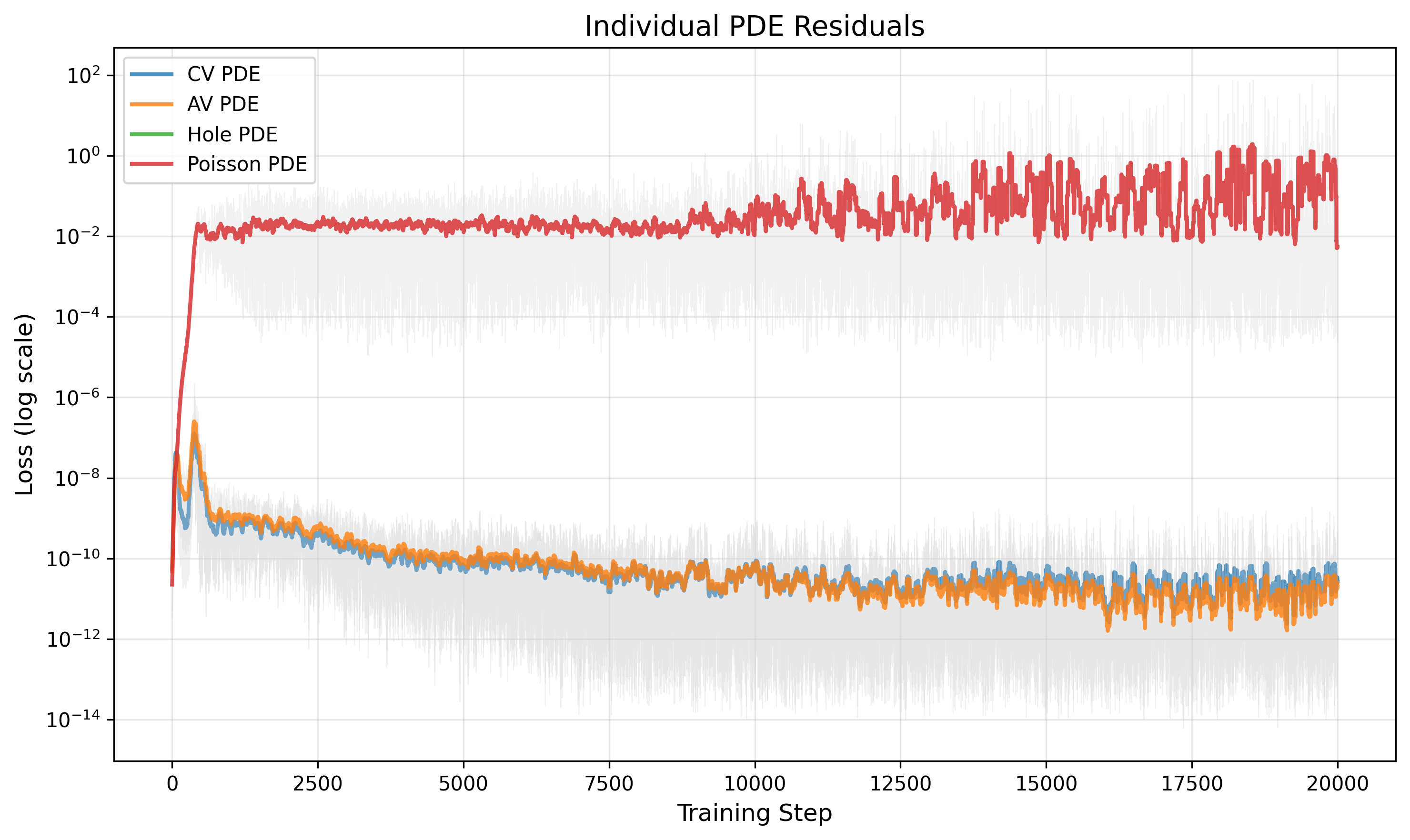}
			\caption{PDE losses, uniform weights}
		\end{subfigure}\hfill
		\begin{subfigure}{0.32\textwidth}
			\includegraphics[width=\textwidth]{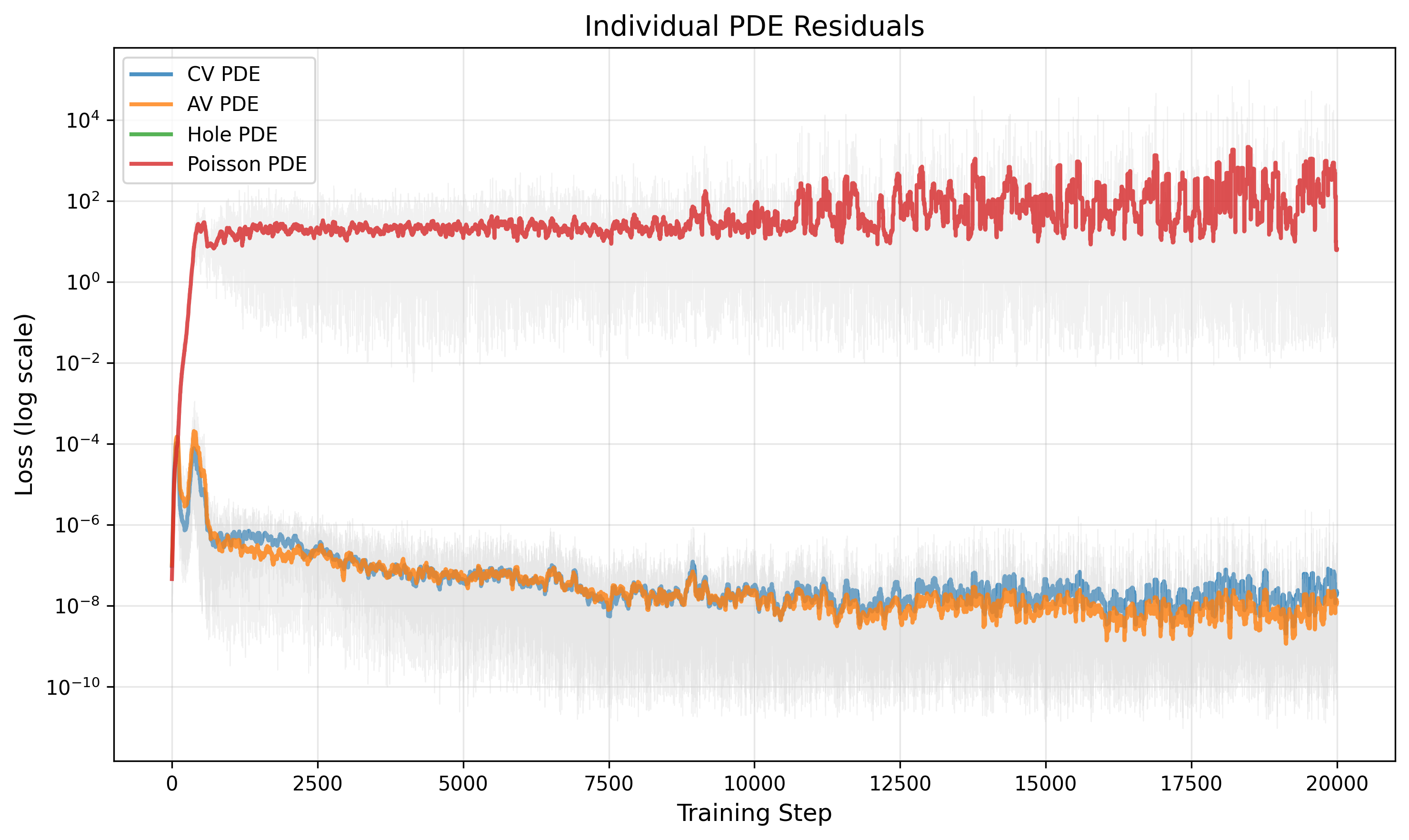}
			\caption{PDE losses, no weights}
		\end{subfigure}\hfill
		\begin{subfigure}{0.32\textwidth}
			\includegraphics[width=\textwidth]{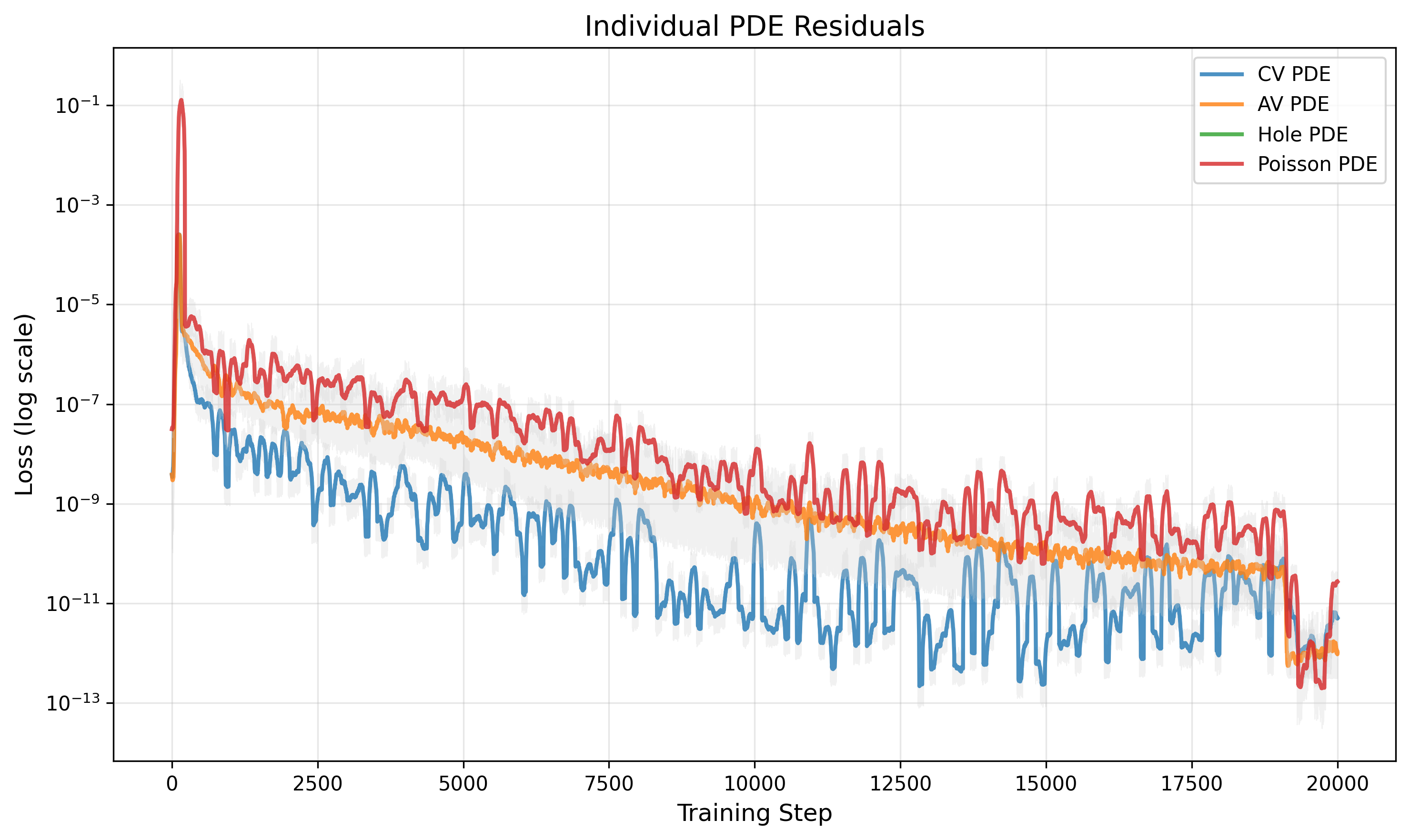}
			\caption{PDE losses, NTK weights}
		\end{subfigure}
		\vspace{0.5em}
		\begin{subfigure}{0.48\textwidth}
			\includegraphics[width=\textwidth]{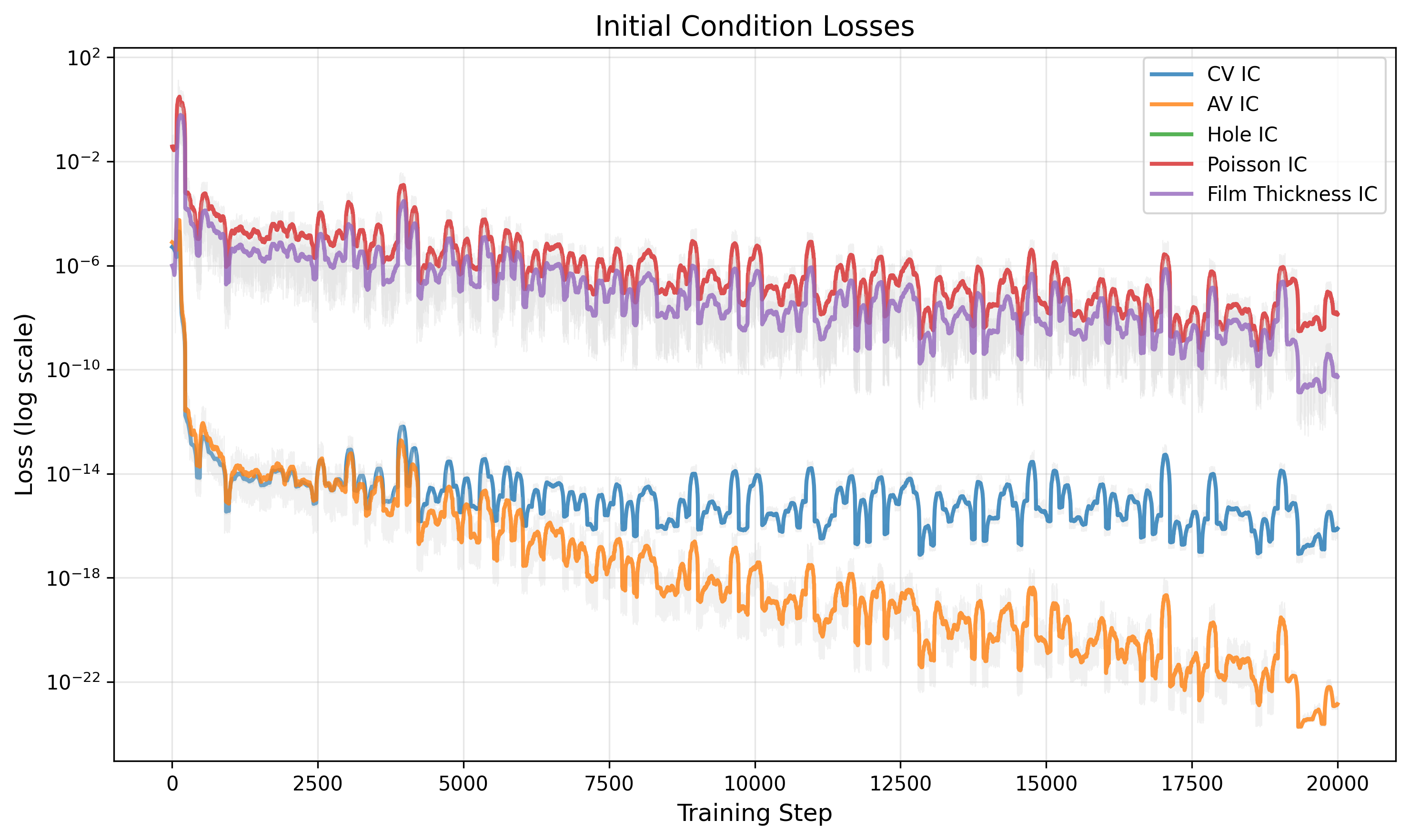}
			\caption{Initial conditions, NTK weights}
		\end{subfigure}\hfill
		\begin{subfigure}{0.48\textwidth}
			\includegraphics[width=\textwidth]{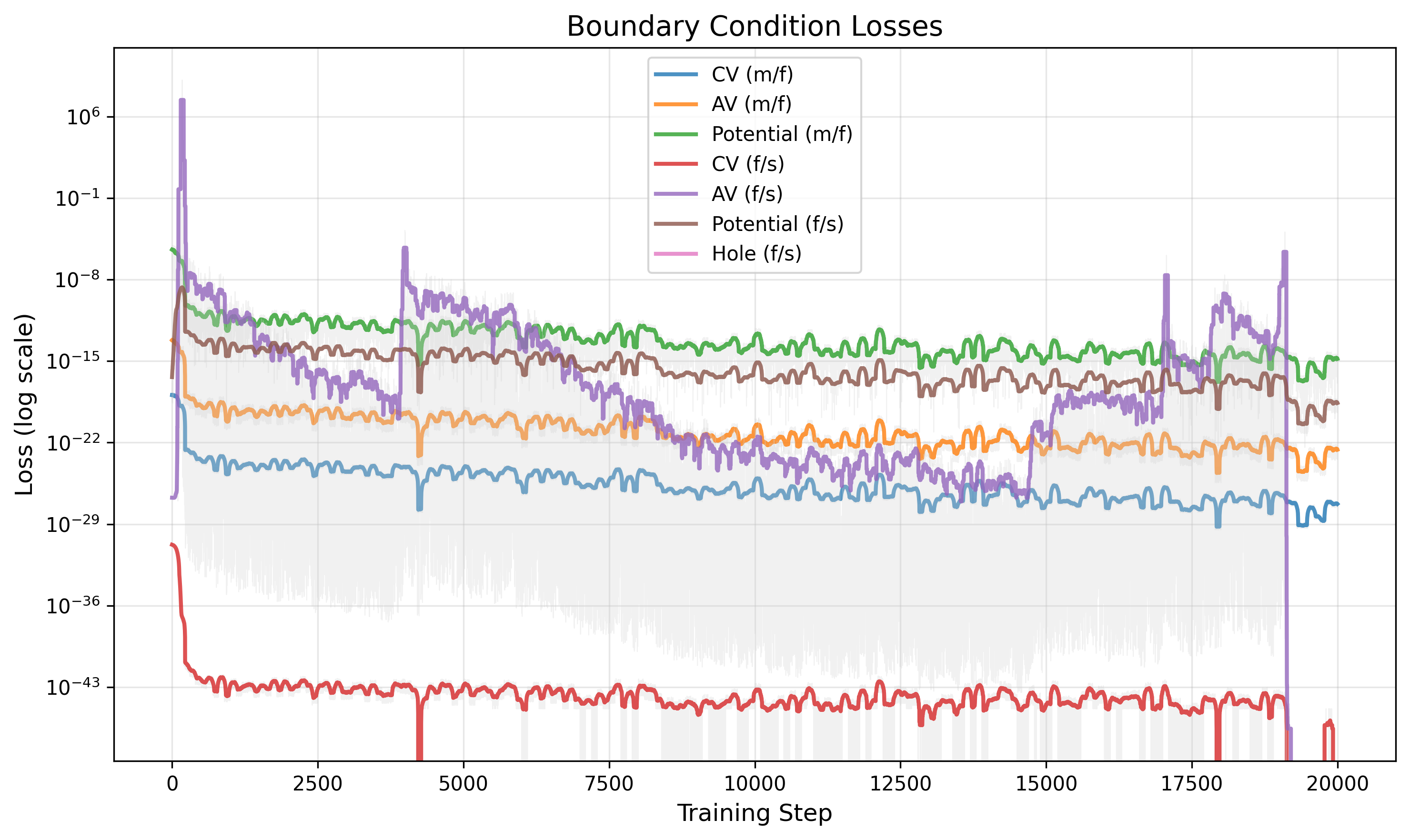}
			\caption{Boundary conditions, NTK weights}
		\end{subfigure}
		\caption{\label{fig:loss_panels}%
			Training loss evolution.  Top row: under uniform weighting (a) and no
			weighting (b), the Poisson PDE residual (red) dominates by four to six
			orders of magnitude, starving the transport equations of gradient
			information; NTK weighting (c) brings all PDE components to within one
			order of magnitude of each other.  Bottom row: NTK weighting also
			produces well-behaved initial conditions (d) and boundary conditions (e),
			though the BC loss remains noisier due to the continuously expanding
			moving-boundary domain.  Solid lines are moving averages; translucent fills
			show raw values.}
	\end{figure*}

	Under uniform weighting, the Poisson loss dominates the transport losses by four
	to six orders of magnitude (Fig.~\ref{fig:loss_panels}(a,b)).  The NTK-weighted
	training brings the Poisson and transport losses to within one order of magnitude
	of each other, and all components show monotonically decreasing trends
	(Fig.~\ref{fig:loss_panels}(c--e)).  Boundary-condition losses remain noisier because the
	moving boundary continuously generates new, previously unseen sampling
	coordinates at each step.  Despite this residual oscillation in the BC loss,
	the converged solution is physically plausible.  Complete loss curves for all
	three weighting configurations are reproduced in the SI (Section~S5) for
	reference.
	
	\subsection{\label{subsec:stiff_bc}Stiff boundary conditions: an open problem}
	
	Some boundary conditions resist convergence regardless of the weighting strategy
	applied.  The NTK approach reduces their assigned weight, which paradoxically
	means the optimizer effectively ignores the hardest constraints.  This is the
	mechanism that leads to qualitatively correct but quantitatively wrong solutions
	in pure PINN mode.
	
	We tested the Augmented Lagrangian (AL) method~\cite{son_enhanced_2023}, which
	reformulates PINN training as a constrained optimisation problem,
	\begin{equation}
		\max_{\bm{\lambda}}\,\min_{\bm{\theta}\in\mathbb{R}^M}
		\left[
		\mathcal{L}_\text{PDE}(\bm{\theta})
		+ \beta\,\|\mathbf{c}(\bm{\theta})\|_2^2
		+ \langle\bm{\lambda},\,\mathbf{c}(\bm{\theta})\rangle
		\right],
		\label{eq:AL}
	\end{equation}
	where $\mathbf{c}(\bm{\theta}) \in \mathbb{R}^{N_c}$ is the vector of
	constraint residuals (boundary and initial conditions), $\beta > 0$ a fixed
	penalty parameter, $\bm{\lambda} \in \mathbb{R}^{N_c}$ the vector of Lagrange
	multipliers, and $\langle\cdot,\cdot\rangle$ the Euclidean inner product.  The
	Lagrange multipliers adapt to enforce the constraints, while the PDE residual
	is minimised in the interior.  Theory guarantees convergence for the Helmholtz,
	viscous Burgers, and Klein--Gordon equations~\cite{son_enhanced_2023}, but
	neither Adam nor L-BFGS converged on the PDM with this formulation.
	
	We also tested staged training in which NTK weighting is used for a warm-up
	phase followed by AL~\cite{bekele_physics-informed_2024}, but this also failed
	to converge.  The
	likely reason is a saddle-point instability: the AL penalty pushes constraints
	toward zero independently of whether the interior loss is satisfied, creating
	dynamics that gradient descent cannot escape from in this strongly coupled system.
	Robust constraint enforcement for stiff multiphysics PINNs is left as an open
	problem.

\begin{figure}[!t]
		\centering
		\includegraphics[width=\columnwidth]{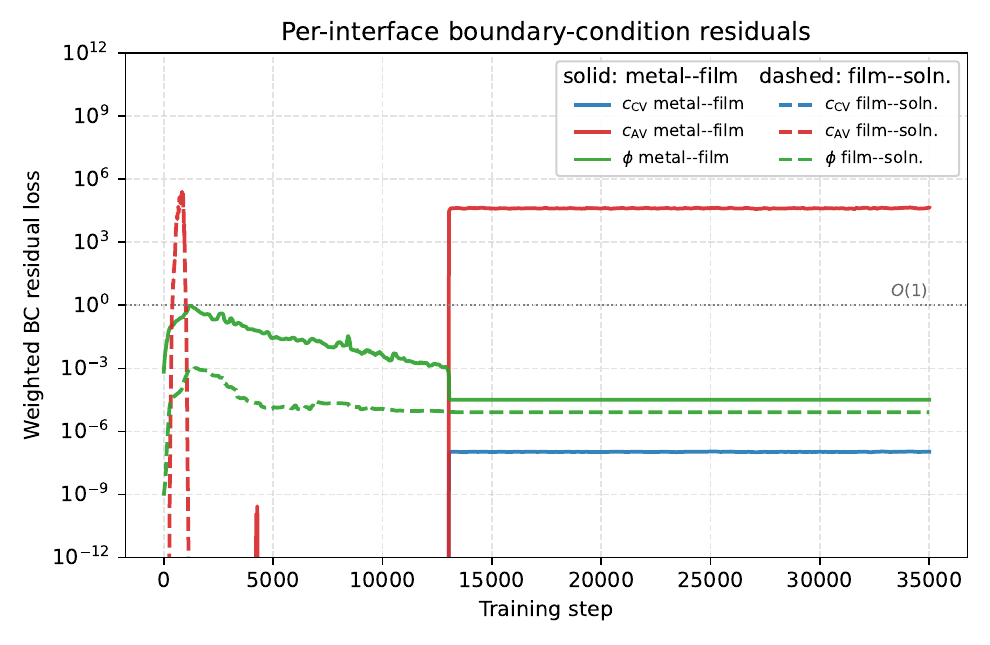}
		\caption{\label{fig:bc_residuals}%
			Per-interface weighted boundary-condition residuals over
			training. Solid lines are metal/film, dashed lines film/solution, for the
			cation-vacancy ($c_\mathrm{CV}$), anion-vacancy ($c_\mathrm{AV}$), and
			potential ($\phi$) constraints. The anion-vacancy residual locks at
			$\mathcal{O}(10^{5})$ after a regime change near step $1.3\times10^{4}$,
			far above all other components.}
	\end{figure}

	To pinpoint which constraint is responsible, we logged the six
	per-interface weighted boundary residuals (the cation-vacancy, anion-vacancy,
	and potential constraints at each of the metal/film and film/solution
	interfaces) throughout training (Fig.~\ref{fig:bc_residuals}). The
	cation-vacancy and potential residuals settle below $10^{-4}$ at both
	interfaces, whereas the anion-vacancy residual undergoes a regime change near
	step $1.3\times10^{4}$ and locks at $\mathcal{O}(10^{5})$, four to five orders
	of magnitude above every other component. This isolates the anion-vacancy
	boundary condition as the single unenforced constraint, consistent with the
	loss-landscape diagnostic below, and as the proximate cause of the
	initialisation-driven divergences seen in Sec.~\ref{subsec:robust}.

	\subsection{Pathological solution convergence: the hybrid remedy}
	
	\label{subsec:patho}
	
	Pure PINN training converges cleanly, but to the wrong solution.  The systematic
	overprediction across all voltages, combined with the qualitatively correct
	exponential growth and voltage dependence, is consistent with
	several non-exclusive explanations, and we present non-uniqueness as a
	hypothesis rather than a settled conclusion. The most direct reading is
	non-uniqueness of the PDM's solution manifold: the coupled Nernst--Planck and
	Poisson equations under exponential Butler--Volmer boundary conditions need not
	admit a unique solution, so the optimiser may settle on a branch that satisfies
	the physics to within tolerance but at the wrong absolute scale. The same
	observations are, however, equally consistent with (i)~incomplete enforcement
	of the stiff anion-vacancy boundary condition, which the boundary-residual
	diagnostic (Fig.~\ref{fig:bc_residuals}) identifies as the dominant unsatisfied
	constraint; (ii)~spectral bias of the Adam-trained network toward low-frequency
	modes early in training; or (iii)~optimisation bias on the residual-loss
	manifold. Distinguishing these rigorously would require an analytical study of
	the PDM solution structure, which is beyond the present scope; either way, the
	single anchor resolves the symptom by selecting the physical branch.
	
	FEM avoids this by implicitly restricting the solution space through spatial
	discretization and physically motivated initialization.  A PINN starting from
	random weights explores a much larger space and can settle on a non-physical
	branch.
	
	Two architectural remedies did not work.  Residual connections smoothed the
	interior loss landscape (Fig.~\ref{fig:loss_landscape}), as expected from Li et
	al.~\cite{li_visualizing_2018}, but this made things worse by allowing the
	optimizer to satisfy interior equations more easily while the boundary constraints
	were neglected.  Enforcing positivity of concentrations via a softmax output
	layer corrupted the gradient information for the film thickness network, yielding
	correct sign but inaccurate magnitudes.
	
	The hybrid approach circumvents the problem by anchoring the solution to the
	correct branch with a single data point.  The reproducibility
	of the converged result across independent random-anchor runs
	(Sec.~\ref{subsec:robust}) confirms that the PINN has already learned the
	physics correctly; one measurement is enough to eliminate the branch
	ambiguity.
	
	\subsection{Loss-landscape diagnostics}
	
\begin{figure*}[!t]
		\centering
		\begin{subfigure}{0.32\textwidth}
			\includegraphics[width=\textwidth]{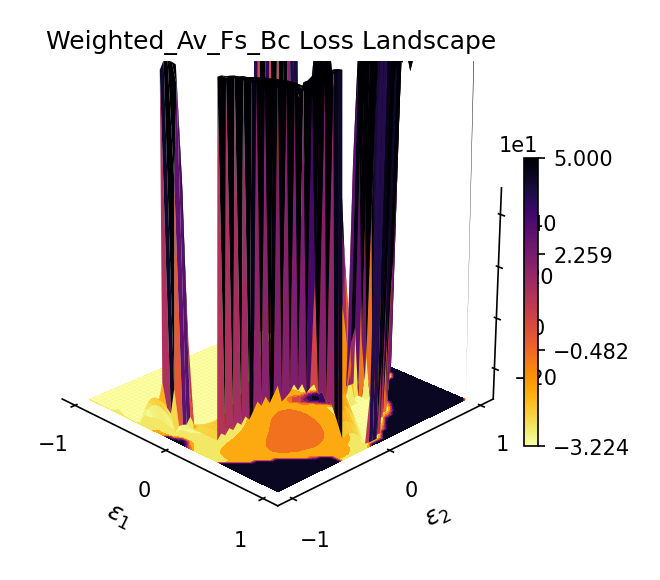}
			\caption{Anion-vacancy BC}
		\end{subfigure}\hfill
		\begin{subfigure}{0.32\textwidth}
			\includegraphics[width=\textwidth]{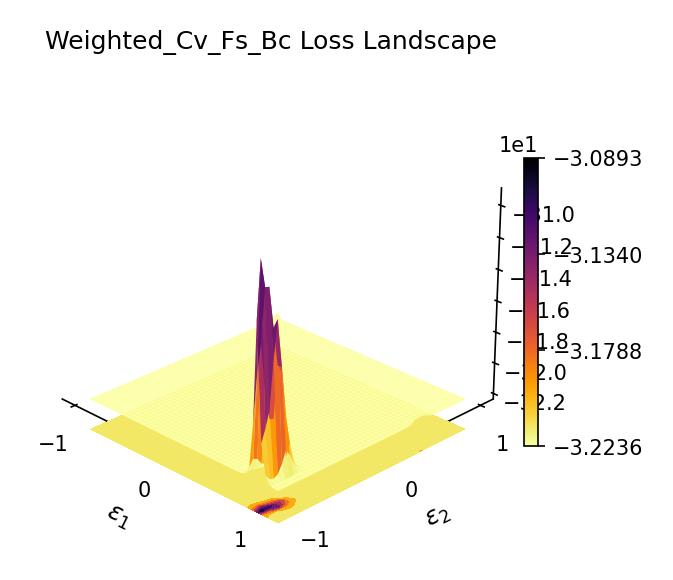}
			\caption{Cation-vacancy BC}
		\end{subfigure}\hfill
		\begin{subfigure}{0.32\textwidth}
			\includegraphics[width=\textwidth]{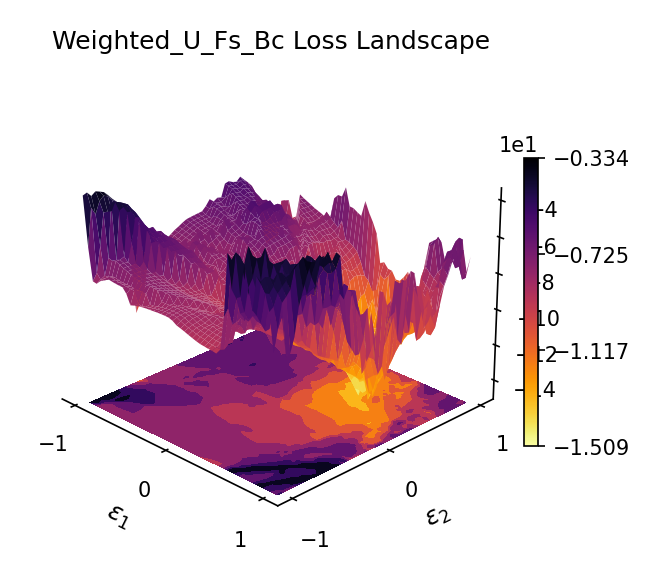}
			\caption{Potential BC}
		\end{subfigure}

		\vspace{0.4em}
		\begin{subfigure}{0.32\textwidth}
			\includegraphics[width=\textwidth]{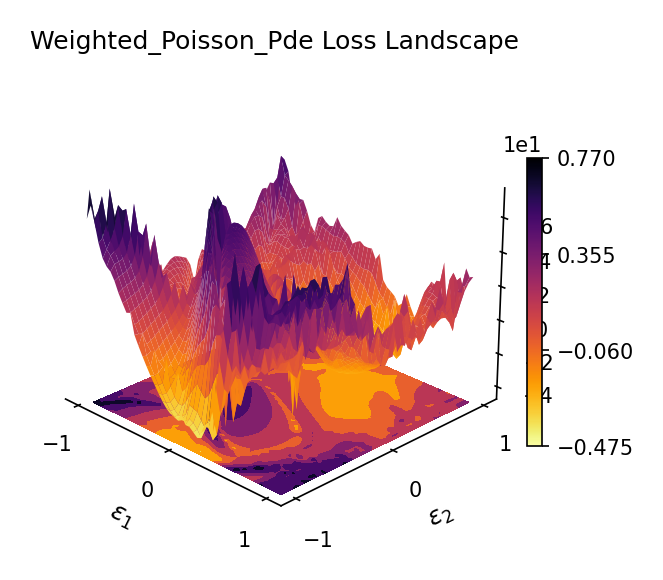}
			\caption{Poisson PDE}
		\end{subfigure}\hfill
		\begin{subfigure}{0.32\textwidth}
			\includegraphics[width=\textwidth]{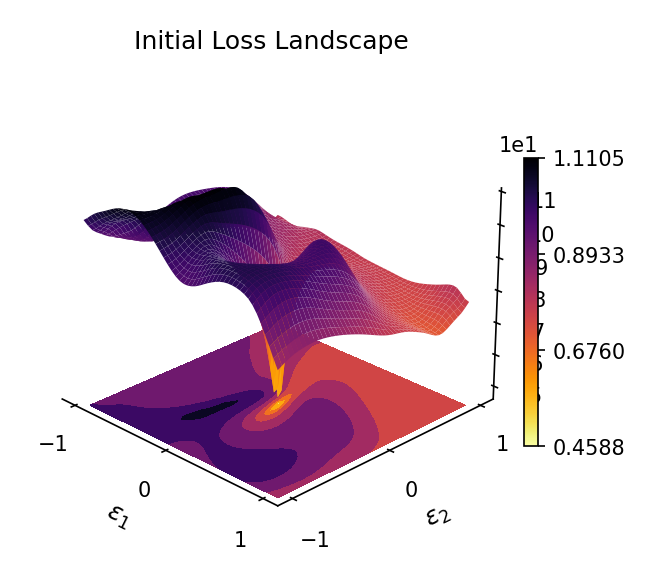}
			\caption{Initial conditions}
		\end{subfigure}\hfill
		\begin{subfigure}{0.32\textwidth}
			\includegraphics[width=\textwidth]{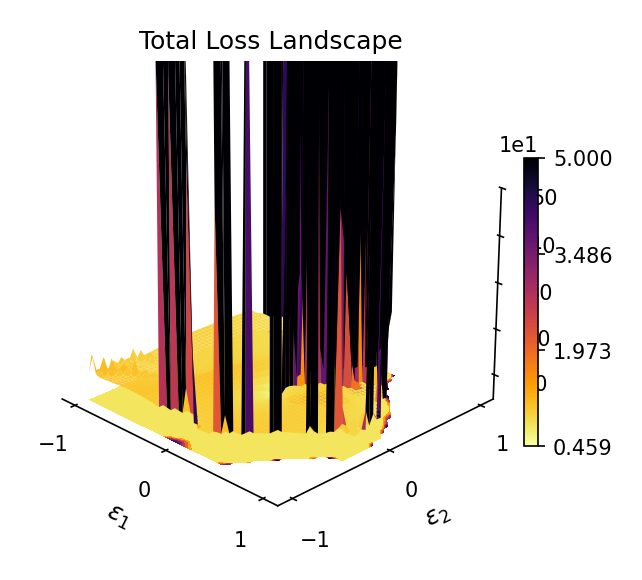}
			\caption{Total loss}
		\end{subfigure}
		\caption{\label{fig:loss_landscape}%
			Loss landscapes around the converged solution, projected onto two
			filter-normalized directions. The anion-vacancy boundary
			condition~(a) is highly corrugated, whereas the other constraints,
			such as the cation-vacancy~(b) and potential~(c) boundary conditions,
			the Poisson residual~(d), and the initial conditions~(e), are
			comparatively smooth. The total landscape~(f) is essentially
			indistinguishable from~(a), identifying the anion-vacancy boundary
			condition as the primary unsatisfied constraint.}
	\end{figure*}

	Loss landscape visualization~\cite{li_visualizing_2018,basir_investigating_2023}
	proved useful for identifying which constraint dominates training difficulties.
	Following Li et al., we project the parameter space onto two random
	filter-normalized directions $\bm{\eta}$ and $\bm{\delta}$, and plot
	\begin{equation}
		f(\alpha,\beta)
		= \mathcal{L}\!\left(\bm{\theta}^* + \alpha\bm{\eta} + \beta\bm{\delta}\right).
		\label{eq:landscape}
	\end{equation}
	Figure~\ref{fig:loss_landscape} shows that the anion vacancy boundary condition
	dominates the total loss landscape, making it the primary target for future
	constraint-enforcement improvements.
	
	\clearpage
	\section{\label{sec:disc}Discussion}
	
	The central finding is that the four failure modes are separable.  The first
	two, scale disparity and loss imbalance, are fully resolved by existing
	techniques: physics-based non-dimensionalization and NTK adaptive weighting.
	Together they bring the pure PINN to qualitatively correct physics.  The
	remaining quantitative error is not a computational pathology but reflects a
	structural property of the PDE system itself.  The coupled Nernst--Planck--Poisson
	equations under exponential Butler--Volmer boundary conditions define a nonlinear
	fixed-point problem; the exponential nonlinearity in the boundary terms does not,
	in general, guarantee a unique solution for arbitrary initial data.  Concretely,
	we observe that different random network initialisations consistently converge to
	solutions with different absolute thickness scales while achieving comparable
	interior residuals, the expected signature of a multiple-branch system.
	Whether this non-uniqueness also manifests in FEM under certain initialisation
	strategies is an open question that warrants analytical study of the PDM's
	solution structure.
	
	The data efficiency of the hybrid approach deserves emphasis.  A purely
	data-driven surrogate for a PDE system of this complexity would require training
	data spanning the full $(x,\,t,\,E)$ space; in practice, generating even a
	coarse coverage with FEM at 10 potentials and 10 time snapshots produces
	$\mathcal{O}(10^2)$ labelled solutions, each requiring a separate COMSOL
	run~\cite{bosing_modeling_2023}.  The hybrid PINN achieves errors
	below $2.2\%$ with a single FEM-validated point, reducing the FEM workload to a one-time spot
	check.  The natural workflow is to use FEM sparingly as a branch selector
	rather than a training set, and then rely on the PINN for rapid parameter-space
	exploration across the full voltage and time range.
	Training the hybrid model is a one-time cost of about
	$47$~minutes on a single GPU, after which the trained networks return $L(t)$ at
	any voltage and time in a single inference pass; reproducing the same coverage
	with FEM requires a separate adaptive-mesh solve for each operating
	condition~\cite{bosing_modeling_2023}, and an inverse or design question
	additionally requires wrapping that solver in an optimisation or adjoint loop.
	A step-by-step checklist
	for applying this workflow to new problems, including common warning signs and
	recommended diagnostic checks, is provided in the SI (Section~S4).
	
	The boundary-condition stiffness problem identified in Sec.~\ref{subsec:stiff_bc}
	is a known challenge across PINN applications~\cite{sophiya_comprehensive_2025,
		toscano_pinns_2024}, and our loss-landscape analysis points specifically to the
	anion vacancy BC as the dominant unsatisfied constraint.  This localisation
	suggests that the path forward is not uniform weight adjustment but structural
	modification of how that particular constraint is enforced.
	Structure-preserving PINNs that embed a Lyapunov function consistent with
	the film-growth thermodynamics~\cite{chu2024structure} could in principle prevent
	the optimizer from exploring regions of weight space that violate energy
	monotonicity, directly reducing the size of the problematic loss basin.
	Kolmogorov--Arnold networks with adaptive activation functions~\cite{zhang_physics-informed_2025}
	offer a complementary route: their per-edge learnable activations can adapt
	locally near the boundary layer where the anion vacancy gradient is steepest,
	potentially matching the stiffness scale without requiring an explicit penalty.
	Energy natural gradient methods~\cite{muller_achieving_2023,guzman-cordero_improving_2025}
	address the same problem from the optimisation side, providing second-order
	curvature information that standard Adam lacks precisely in the flat-loss
	regions around stiff constraints.
	
	Regarding the PINN's limitation to fixed geometry: the approach as implemented
	handles one-dimensional moving boundaries through the film-growth ODE.
	Extension to two- or three-dimensional geometries would require re-training for
	each geometry change, as is standard for PINNs~\cite{toscano_pinns_2024}.
	Finite Basis PINNs~\cite{moseley_finite_2023} or domain-decomposition approaches
	could address this limitation and constitute a natural next step.
	
	The methodology developed here (non-dimensionalization, NTK weighting, hybrid
	anchoring, and loss-landscape diagnostics) is not specific to the PDM.  The
	four failure modes recur whenever a PINN couples stiff transport to nonlinear
	interface kinetics.  In dendritic solidification, for instance, the
	Nernst--Planck equation for solute is replaced by a heat equation, but the
	stiffness structure is identical: the solid/liquid interface moves on a time
	scale set by latent heat release that is orders of magnitude shorter than bulk
	diffusion, producing the same scale-disparity failure that dimensional PDM
	training exhibits.  Reactive transport in solid electrolytes presents a
	two-failure-mode problem: Butler--Volmer exchange at grain boundaries provides
	the non-unique branch structure, while the wide range of ionic conductivities
	across phases produces the loss imbalance.  In both cases the hybrid anchoring
	strategy, which uses a single validated reference solution to select the physical
	branch, transfers without modification.
	
	\section{\label{sec:conc}Conclusions}
	
	We have shown that physics-informed neural networks can solve
	the Point Defect Model of passive-film growth and, more valuably, recover its
	kinetic parameters from sparse data, a problem central to predicting material
	degradation and one for which conventional approaches lean on specialised FEM
	solvers and costly experiments. The PDM is difficult for reasons common to
	stiff multiphysics systems rather than peculiar to it: widely separated scales,
	stiff boundary conditions, and convergence to non-physical solution branches. We
	handled the first with physics-based non-dimensionalisation (stable simulation
	from about one hour to 250~hours), the second with NTK adaptive weighting (a
	four-to-six-order loss imbalance compressed to roughly one), and the branch
	ambiguity with a single validated anchor that brings film-thickness error below
	$2.2\%$ at all five potentials; stiff boundary-condition enforcement resisted
	every weighting scheme and remains open, with a boundary-residual and
	loss-landscape analysis localising it to the anion-vacancy condition. The result
	is robust because the anchor is resampled at every step, accuracy saturates
	beyond about ten anchors, and the physics loss tolerates roughly $5\%$
	measurement noise. The practical payoff is that this anchor can be a single
	experiment: from sparse measurements the framework reconstructs the full coupled
	solution and infers unknown kinetic constants, with recoverability tracking
	stiffness (the boundary-stiff rate constant is identifiable, the weakly-coupled
	interior coefficient is not). By turning a few measurements into full fields and
	calibrated kinetics, it reduces the experimental and computational burden of
	studying passive-film growth, and the same recipe transfers to other stiff
	transport--reaction systems with moving boundaries.

	\section*{Supplementary Material}
	See the supplementary material for the complete interfacial reaction schemes,
	the characteristic scales and full non-dimensionalisation of the governing
	equations, a table of all model parameters with values and units, practical
	guidelines for applying the framework to new stiff multiphysics problems,
	additional training-loss curves, the random-anchor robustness analysis, and the
	inverse-problem setup.

	\begin{acknowledgments}
		This work was supported by the NSERC--CNSC Small Modular Reactors Research Grant
		Initiative (ALLRP~580475-2022).  This work was made possible by the facilities of the Shared Hierarchical 
		Academic Research Computing Network (SHARCNET:www.sharcnet.ca) and Digital Research Alliance of Canada (https://alliancecan.ca/en).
	\end{acknowledgments}

	\section*{Author Contributions}
	\textbf{Mohid Farooqi}: Conceptualization (equal); Software (lead); Formal
	analysis (lead); Investigation (lead); Visualization (lead); Writing -- original
	draft (lead); Writing -- review \& editing (equal). \textbf{Ingmar B\"{o}sing}:
	Resources (lead); Data curation (lead); Writing -- review \& editing
	(supporting). \textbf{Conrard Giresse Tetsassi Feugmo}: Conceptualization
	(lead); Supervision (lead); Funding acquisition (lead); Software (supporting);
	Writing -- original draft (supporting); Writing -- review \& editing (equal).

	\section*{Data Availability Statement}

	The Python implementation of the PINNACLE training framework, all model
	parameters, and scripts to reproduce all figures in this paper are available at
	\url{https://github.com/Feugmo-Group/PINNACLE} under an open-source MIT licence.
	FEM reference data are included in the repository in tabular form.
	Scripts to reproduce the ablation, robustness,
	data-efficiency, noise, and inverse-problem experiments reported in
	Sec.~\ref{sec:results} are provided in the \texttt{scripts/} directory, and
	the Hydra configuration used for all reported runs is provided under
	\texttt{conf/}.
	
	\bibliography{references}

@article{bosing_modeling_2023,
	title = {Modeling electrochemical oxide film growth—passive and transpassive behavior of iron electrodes in halide-free solution},
	volume = {7},
	issn = {2397-2106},
	url = {https://www.nature.com/articles/s41529-023-00369-y},
	doi = {10.1038/s41529-023-00369-y},
number = {1},
	urldate = {2025-05-15},
	journal = {npj Materials Degradation},
	author = {Bösing, Ingmar},
	month = jun,
	year = {2023},
	pages = {53},
}

@article{macdonald_history_2011,
	title = {The history of the {Point} {Defect} {Model} for the passive state: {A} brief review of film growth aspects},
	volume = {56},
	copyright = {https://www.elsevier.com/tdm/userlicense/1.0/},
	issn = {00134686},
	shorttitle = {The history of the {Point} {Defect} {Model} for the passive state},
	url = {https://linkinghub.elsevier.com/retrieve/pii/S001346861001515X},
	doi = {10.1016/j.electacta.2010.11.005},
number = {4},
	urldate = {2025-05-15},
	journal = {Electrochimica Acta},
	author = {Macdonald, Digby D.},
	month = jan,
	year = {2011},
	pages = {1761--1772},
}

@article{sun_physics-informed_2024,
	title = {A physics-informed neural network framework for multi-physics coupling microfluidic problems},
	volume = {284},
	issn = {00457930},
	url = {https://linkinghub.elsevier.com/retrieve/pii/S0045793024002524},
	doi = {10.1016/j.compfluid.2024.106421},
urldate = {2025-05-20},
	journal = {Computers \& Fluids},
	author = {Sun, Runze and Jeong, Hyogu and Zhao, Jiachen and Gou, Yixing and Sauret, Emilie and Li, Zirui and Gu, Yuantong},
	month = nov,
	year = {2024},
	pages = {106421},
}

@article{li_tutorials_2024,
	title = {Tutorials: {Physics}-informed machine learning methods of computing {1D} phase-field models},
	volume = {2},
	issn = {2770-9019},
	shorttitle = {Tutorials},
	url = {https://pubs.aip.org/aml/article/2/3/031101/3308986/Tutorials-Physics-informed-machine-learning},
	doi = {10.1063/5.0205159},
number = {3},
	urldate = {2025-05-20},
	journal = {APL Machine Learning},
	author = {Li, Wei and Fang, Ruqing and Jiao, Junning and Vassilakis, Georgios N. and Zhu, Juner},
	month = sep,
	year = {2024},
	pages = {031101},
}

@misc{kathane_physics_2024,
	title = {A {Physics} {Informed} {Neural} {Network} ({PINN}) {Methodology} for {Coupled} {Moving} {Boundary} {PDEs}},
	url = {http://arxiv.org/abs/2409.10910},
	doi = {10.48550/arXiv.2409.10910},
urldate = {2025-05-22},
	publisher = {arXiv},
	author = {Kathane, Shivprasad and Karagadde, Shyamprasad},
	month = sep,
	year = {2024},
	note = {arXiv:2409.10910 [cs]},
}

@misc{toscano_pinns_2024,
	title = {From {PINNs} to {PIKANs}: {Recent} {Advances} in {Physics}-{Informed} {Machine} {Learning}},
	shorttitle = {From {PINNs} to {PIKANs}},
	url = {http://arxiv.org/abs/2410.13228},
	doi = {10.48550/arXiv.2410.13228},
urldate = {2025-06-17},
	publisher = {arXiv},
	author = {Toscano, Juan Diego and Oommen, Vivek and Varghese, Alan John and Zou, Zongren and Daryakenari, Nazanin Ahmadi and Wu, Chenxi and Karniadakis, George Em},
	month = oct,
	year = {2024},
	note = {arXiv:2410.13228 [cs]},
}

@misc{son_enhanced_2023,
	title = {Enhanced {Physics}-{Informed} {Neural} {Networks} with {Augmented} {Lagrangian} {Relaxation} {Method} ({AL}-{PINNs})},
	url = {http://arxiv.org/abs/2205.01059},
	doi = {10.48550/arXiv.2205.01059},
urldate = {2025-07-16},
	publisher = {arXiv},
	author = {Son, Hwijae and Cho, Sung Woong and Hwang, Hyung Ju},
	month = may,
	year = {2023},
	note = {arXiv:2205.01059 [cs]},
}

@article{moseley_finite_2023,
	title = {Finite {Basis} {Physics}-{Informed} {Neural} {Networks} ({FBPINNs}): a scalable domain decomposition approach for solving differential equations},
	volume = {49},
	issn = {1019-7168, 1572-9044},
	shorttitle = {Finite {Basis} {Physics}-{Informed} {Neural} {Networks} ({FBPINNs})},
	url = {http://arxiv.org/abs/2107.07871},
	doi = {10.1007/s10444-023-10065-9},
number = {4},
	urldate = {2025-07-15},
	journal = {Advances in Computational Mathematics},
	author = {Moseley, Ben and Markham, Andrew and Nissen-Meyer, Tarje},
	month = aug,
	year = {2023},
	note = {arXiv:2107.07871 [physics]},
}

@misc{li_visualizing_2018,
	title = {Visualizing the {Loss} {Landscape} of {Neural} {Nets}},
	url = {http://arxiv.org/abs/1712.09913},
	doi = {10.48550/arXiv.1712.09913},
urldate = {2025-07-10},
	publisher = {arXiv},
	author = {Li, Hao and Xu, Zheng and Taylor, Gavin and Studer, Christoph and Goldstein, Tom},
	month = nov,
	year = {2018},
	note = {arXiv:1712.09913 [cs]},
}

@article{basir_investigating_2023,
	title = {Investigating and {Mitigating} {Failure} {Modes} in {Physics}-informed {Neural} {Networks} ({PINNs})},
	volume = {33},
	issn = {1991-7120, 1815-2406},
	url = {http://arxiv.org/abs/2209.09988},
	doi = {10.4208/cicp.OA-2022-0239},
number = {5},
	urldate = {2025-06-20},
	journal = {Communications in Computational Physics},
	author = {Basir, Shamsulhaq},
	month = jan,
	year = {2023},
	note = {arXiv:2209.09988 [cs]},
	pages = {1240--1269},
}

@article{zhang_physics-informed_2025,
	title = {Physics-informed neural networks with hybrid {Kolmogorov}-{Arnold} network and augmented {Lagrangian} function for solving partial differential equations},
	volume = {15},
	copyright = {2025 The Author(s)},
	issn = {2045-2322},
	url = {https://www.nature.com/articles/s41598-025-92900-1},
	doi = {10.1038/s41598-025-92900-1},
number = {1},
	urldate = {2025-06-20},
	journal = {Scientific Reports},
	author = {Zhang, Zhaoyang and Wang, Qingwang and Zhang, Yinxing and Shen, Tao and Zhang, Weiyi},
	month = mar,
	year = {2025},
	note = {Publisher: Nature Publishing Group},
	pages = {10523},
}

@misc{jacot_neural_2020,
	title = {Neural {Tangent} {Kernel}: {Convergence} and {Generalization} in {Neural} {Networks}},
	shorttitle = {Neural {Tangent} {Kernel}},
	url = {http://arxiv.org/abs/1806.07572},
	doi = {10.48550/arXiv.1806.07572},
urldate = {2025-06-17},
	publisher = {arXiv},
	author = {Jacot, Arthur and Gabriel, Franck and Hongler, Clément},
	month = feb,
	year = {2020},
	note = {arXiv:1806.07572 [cs]},
}

@misc{al-safwan_is_2021,
	title = {Is it time to swish? {Comparing} activation functions in solving the {Helmholtz} equation using physics-informed neural networks},
	shorttitle = {Is it time to swish?},
	url = {http://arxiv.org/abs/2110.07721},
	doi = {10.48550/arXiv.2110.07721},
urldate = {2025-08-01},
	publisher = {arXiv},
	author = {Al-Safwan, Ali and Song, Chao and Waheed, Umair bin},
	month = oct,
	year = {2021},
	note = {arXiv:2110.07721 [physics]},
}

@misc{muller_achieving_2023,
	title = {Achieving {High} {Accuracy} with {PINNs} via {Energy} {Natural} {Gradients}},
	url = {http://arxiv.org/abs/2302.13163},
	doi = {10.48550/arXiv.2302.13163},
urldate = {2025-08-05},
	publisher = {arXiv},
	author = {Müller, Johannes and Zeinhofer, Marius},
	month = aug,
	year = {2023},
	note = {arXiv:2302.13163 [cs]},
}

@misc{guzman-cordero_improving_2025,
	title = {Improving {Energy} {Natural} {Gradient} {Descent} through {Woodbury}, {Momentum}, and {Randomization}},
	url = {http://arxiv.org/abs/2505.12149},
	doi = {10.48550/arXiv.2505.12149},
urldate = {2025-08-05},
	publisher = {arXiv},
	author = {Guzmán-Cordero, Andrés and Dangel, Felix and Goldshlager, Gil and Zeinhofer, Marius},
	month = may,
	year = {2025},
	note = {arXiv:2505.12149 [cs]},
}

@article{bekele_physics-informed_2024,
	title = {{PHYSICS}-{INFORMED} {NEURAL} {NETWORKS} {WITH} {CURRICULUM} {TRAINING} {FOR} {POROELASTIC} {FLOW} {AND} {DEFORMATION} {PROCESSES}},
journal = {Computers and Geotechnics},
	author = {Bekele, Yared W},
	year = {2024},
}

@incollection{singh_harsh_2016,
    title = {Harsh {Environment} {Materials}},
    isbn = {978-0-12-803581-8},
    url = {https://www.sciencedirect.com/science/article/pii/B9780128035818092535},
    urldate = {2025-08-10},
    booktitle = {Reference {Module} in {Materials} {Science} and {Materials} {Engineering}},
    publisher = {Elsevier},
    author = {Singh, T. and Kohn, E.},
    month = jan,
    year = {2016},
    doi = {10.1016/B978-0-12-803581-8.09253-5},
}

@article{raissi_physics-informed_2019,
    title = {Physics-informed neural networks: {A} deep learning framework for solving forward and inverse problems involving nonlinear partial differential equations},
    volume = {378},
    issn = {0021-9991},
    shorttitle = {Physics-informed neural networks},
    url = {https://www.sciencedirect.com/science/article/pii/S0021999118307125},
    doi = {10.1016/j.jcp.2018.10.045},
    urldate = {2025-08-10},
    journal = {Journal of Computational Physics},
    author = {Raissi, M. and Perdikaris, P. and Karniadakis, G. E.},
    month = feb,
    year = {2019},
    pages = {686--707},
}

@Article{chen_hydrodynamic_2022,
author ="Chen, Haotian and Kätelhön, Enno and Compton, Richard G.",
title  ="The application of physics-informed neural networks to hydrodynamic voltammetry",
journal  ="Analyst",
year  ="2022",
volume  ="147",
issue  ="9",
pages  ="1881-1891",
publisher  ="The Royal Society of Chemistry",
doi  ="10.1039/D2AN00456A",
url  ="http://dx.doi.org/10.1039/D2AN00456A"}

@article{chen2022critical-abc, 
  year     = {2022}, 
  title    = {A Critical Evaluation of Using Physics-Informed Neural Networks for Simulating Voltammetry: Strengths, Weaknesses and Best Practices}, 
  author   = {Chen, Haotian and Batchelor-{McAuley}, Christopher and Kätelhön, Enno and Elliott, Joseph and Compton, Richard G.}, 
  journal  = {Journal of Electroanalytical Chemistry}, 
  issn     = {1572-6657}, 
  doi      = {10.1016/j.jelechem.2022.116918}, 
  pages    = {116918}, 
  volume   = {925}
}

@article{chen2022predicting-abc, 
  year     = {2022}, 
  title    = {Predicting Voltammetry Using Physics-Informed Neural Networks}, 
  author   = {Chen, Haotian and Kätelhön, Enno and Compton, Richard G}, 
  journal  = {The Journal of Physical Chemistry Letters}, 
  issn     = {1948-7185}, 
  doi      = {10.1021/acs.jpclett.1c04054}, 
  pmid     = {35007069}, 
  pmcid    = {{PMC}9084599}, 
  pages    = {536--543}, 
  number   = {2}, 
  volume   = {13}
}

@article{Iannuzzi2022,
  author = {Iannuzzi, M. and Frankel, G. S.},
  title = {The carbon footprint of steel corrosion},
  journal = {npj Materials Degradation},
  volume = {6},
  pages = {101},
  year = {2022},
  doi = {10.1038/s41529-022-00318-1},
  url = {https://doi.org/10.1038/s41529-022-00318-1}
}

@article{ALAMIERY2024102672,
title = {Sustainable corrosion Inhibitors: A key step towards environmentally responsible corrosion control},
journal = {Ain Shams Engineering Journal},
volume = {15},
number = {5},
pages = {102672},
year = {2024},
issn = {2090-4479},
doi = {https://doi.org/10.1016/j.asej.2024.102672},
url = {https://www.sciencedirect.com/science/article/pii/S2090447924000479},
author = {Ahmed Al-Amiery and Wan Nor Roslam {Wan Isahak} and Waleed Khalid Al-Azzawi}
}

@article{fu2021catalytic-646, 
  year     = {2021}, 
  title    = {Catalytic open-circuit passivation by thin metal oxide films of p-Si anodes in aqueous alkaline electrolytes}, 
  author   = {Fu, Harold J. and Buabthong, Pakpoom and Ifkovits, Zachary Philip and Yu, Weilai and Brunschwig, Bruce S. and Lewis, Nathan S.}, 
  journal  = {Energy \& Environmental Science}, 
  issn     = {1754-5692}, 
  doi      = {10.1039/d1ee03040j}, 
  pages    = {334--345}, 
  number   = {1}, 
  volume   = {15}
}

@article{ma2019origin-e88, 
  year     = {2019}, 
  title    = {Origin of nanoscale heterogeneity in the surface oxide film protecting stainless steel against corrosion}, 
  author   = {Ma, Li and Wiame, Frédéric and Maurice, Vincent and Marcus, Philippe}, 
  journal  = {npj Materials Degradation}, 
  doi      = {10.1038/s41529-019-0091-4}, 
  pages    = {29}, 
  number   = {1}, 
  volume   = {3}
}

@article{maurice2018current-96f, 
  year     = {2018}, 
  title    = {Current developments of nanoscale insight into corrosion protection by passive oxide films}, 
  author   = {Maurice, Vincent and Marcus, Philippe}, 
  journal  = {Current Opinion in Solid State and Materials Science}, 
  issn     = {1359-0286}, 
  doi      = {10.1016/j.cossms.2018.05.004}, 
  eprint   = {1807.10689}, 
  pages    = {156--167}, 
  number   = {4}, 
  volume   = {22}
}

@article{Seyeux.2013, 
year = {2013}, 
title = {{Oxide Film Growth Kinetics on Metals and Alloys: I. Physical Model}}, 
author = {Seyeux, Antoine and Maurice, Vincent and Marcus, Philippe}, 
journal = {Journal of The Electrochemical Society}, 
issn = {0013-4651}, 
doi = {10.1149/2.036306jes}, 
pages = {C189--C196}, 
number = {6}, 
volume = {160}
}

@article{Leistner.2013, 
year = {2013}, 
title = {{Oxide Film Growth Kinetics on Metals and Alloys: II. Numerical Simulation of Transient Behavior}}, 
author = {Leistner, Kirsten and Toulemonde, Charles and Diawara, Boubakar and Seyeux, Antoine and Marcus, Philippe}, 
journal = {Journal of The Electrochemical Society}, 
issn = {0013-4651}, 
doi = {10.1149/2.037306jes}, 
pages = {C197--C205}, 
number = {6}, 
volume = {160}
}

@article{Li_PDM, 
year = {2020}, 
title = {{Point defect model for the corrosion of steels in supercritical water: Part I, film growth kinetics}}, 
author = {Li, Yanhui and Macdonald, Digby D. and Yang, Jie and Qiu, Jie and Wang, Shuzhong}, 
journal = {Corrosion Science}, 
issn = {0010-938X}, 
doi = {10.1016/j.corsci.2019.108280}, 
pages = {108280}, 
volume = {163}
}

@article{Kolotinskii_2023, 
year = {2023}, 
title = {{Point Defect Model for the kinetics of oxide film growth on the surface of T91 steel in contact with lead–bismuth eutectic}}, 
author = {Kolotinskii, D.A. and Nikolaev, V.S. and Stegailov, V.V. and Timofeev, A.V.}, 
journal = {Corrosion Science}, 
issn = {0010-938X}, 
doi = {10.1016/j.corsci.2022.110829}, 
pages = {110829}, 
volume = {211}
}

@article{alexiadis2025modeling-ef0, 
  year     = {2025}, 
  title    = {Modeling and simulation of passive film formation and breakdown in chloride ion containing electrolytes – A Point Defect Model extension}, 
  author   = {Alexiadis, Nikos and Fuchs, Alexander and Troßmann, Torsten and Bösing, Ingmar and Thöming, Jorg and Mantia, Fabio La}, 
  journal  = {Corrosion Science}, 
  issn     = {0010-938X}, 
  doi      = {10.1016/j.corsci.2025.113166}, 
  pages    = {113166}, 
  volume   = {256}
}

@article{engelhardt2024estimation-16a, 
  year     = {2024}, 
  title    = {Estimation of Some Parameters in the Point Defect Model ({PDM}) for the Passivity of Metals}, 
  author   = {Engelhardt, George R. and Chen, Dihao and Dong, Chaofang and Macdonald, Digby D.}, 
  journal  = {Journal of The Electrochemical Society}, 
  issn     = {0013-4651}, 
  doi      = {10.1149/1945-7111/ad318e}, 
  pages    = {031503}, 
  number   = {3}, 
  volume   = {171}
}

@article{bataillon2012numerical-9fe, 
  year     = {2012}, 
  title    = {Numerical methods for the simulation of a corrosion model with moving oxide layer}, 
  author   = {Bataillon, C. and Bouchon, F. and Chainais-Hillairet, C. and Fuhrmann, J. and Hoarau, E. and Touzani, R.}, 
  journal  = {Journal of Computational Physics}, 
  issn     = {0021-9991}, 
  doi      = {10.1016/j.jcp.2012.06.005}, 
  pages    = {6213--6231}, 
  number   = {18}, 
  volume   = {231}
}

@article{chen2025pfpinns-abc, 
  year     = {2025}, 
  title    = {{PF}-{PINNs}: Physics-informed neural networks for solving coupled Allen-Cahn and Cahn-Hilliard phase field equations}, 
  author   = {Chen, Nanxi and Lucarini, Sergio and Ma, Rujin and Chen, Airong and Cui, Chuanjie}, 
  journal  = {Journal of Computational Physics}, 
  issn     = {0021-9991}, 
  doi      = {10.1016/j.jcp.2025.113843}, 
  pages    = {113843}, 
  volume   = {529}
}

@article{sun2022point-49d, 
  year     = {2022}, 
  title    = {Point defect model for passivity breakdown on hyper-duplex stainless steel 2707 in solutions containing bromide at different temperatures}, 
  author   = {Sun, Li and Zhao, Tianyu and Qiu, Jie and Sun, Yangting and Li, Kuijiao and Zheng, Haibing and Jiang, Yiming and Li, Yanhui and Li, Jin and Li, Weihua and Macdonald, Digby D.}, 
  journal  = {Corrosion Science}, 
  issn     = {0010-938X}, 
  doi      = {10.1016/j.corsci.2021.109959}, 
  pages    = {109959}, 
  volume   = {194}
}

@article{malekjani_comparative_2025, 
  year     = {2025}, 
  title    = {A comparative study of dimensional and non-dimensional inputs in physics-informed and data-driven neural networks for single-droplet evaporation}, 
  author   = {Malekjani, Narjes and Kharaghani, Abdolreza and Tsotsas, Evangelos}, 
  journal  = {Chemical Engineering Science}, 
  issn     = {0009-2509}, 
  doi      = {10.1016/j.ces.2025.121214}, 
  pages    = {121214}, 
  volume   = {306}
}

@misc{sophiya_comprehensive_2025,
    title = {A comprehensive analysis of {PINNs}: {Variants}, {Applications}, and {Challenges}},
    shorttitle = {A comprehensive analysis of {PINNs}},
    url = {http://arxiv.org/abs/2505.22761},
    doi = {10.48550/arXiv.2505.22761},
urldate = {2025-08-22},
    publisher = {arXiv},
    author = {Sophiya, Afila Ajithkumar and Nair, Akarsh K. and Maleki, Sepehr and Krishnababu, Senthil K.},
    month = may,
    year = {2025},
    note = {arXiv:2505.22761 [cs]},
}

@inproceedings{chu2024structure,
  title={Structure-Preserving Physics-Informed Neural Networks with Energy or Lyapunov Structure},
  author={Chu, Haoyu and Miyatake, Yuto and Cui, Wenjun and Wei, Shikui and Furihata, Daisuke},
  booktitle={Proceedings of the Thirty-Third International Joint Conference on Artificial Intelligence},
  pages={3872--3880},
  year={2024},
  doi={10.24963/ijcai.2024/428},
  url={https://doi.org/10.24963/ijcai.2024/428}
}

@article{raissi2020hidden,
  title = {Hidden fluid mechanics: Learning velocity and pressure fields from flow visualizations},
  author = {Raissi, Maziar and Yazdani, Alireza and Karniadakis, George Em},
  journal = {Science},
  volume = {367},
  number = {6481},
  pages = {1026--1030},
  year = {2020},
  publisher = {American Association for the Advancement of Science},
  doi = {10.1126/science.aaw4741},
}

@article{chen2021scarce,
  title   = {Physics-informed learning of governing equations from scarce data},
  author  = {Chen, Zhao and Liu, Yang and Sun, Hao},
  journal = {Nature Communications},
  volume  = {12},
  pages   = {6136},
  year    = {2021},
  doi     = {10.1038/s41467-021-26434-1},
}

@article{travnikova2025quantifying,
  title = {Quantifying data needs in surrogate modeling for flow fields in two-dimensional stirred tanks with physics-informed neural networks},
  author = {Tr{\'a}vn{\'\i}kov{\'a}, V. and Von Lieres, E. and Behr, M.},
  journal = {Physics of Fluids},
  year = {2025},
  publisher = {AIP Publishing},
}

@article{bhatnagar2024pinn,
  title = {Physics Informed Neural Networks for Modeling of {3D} Flow--Thermal Problems with Sparse Domain Data},
  author = {Bhatnagar, S. and Comerford, A. and Banaeizadeh, A.},
  journal = {Journal of Machine Learning for Modeling and Computing},
  volume = {5},
  number = {1},
  pages = {39--67},
  year = {2024},
}

@article{cuomo2022scientific,
  title = {Scientific machine learning through physics-informed neural networks: Where we are and what's next},
  author = {Cuomo, Salvatore and Di Cola, Vincenzo Schiano and Giampaolo, Fabio and Rozza, Gianluigi and Raissi, Maziar and Piccialli, Francesco},
  journal = {Journal of Scientific Computing},
  volume = {92},
  number = {3},
  pages = {88},
  year = {2022},
  publisher = {Springer},
  doi = {10.1007/s10915-022-01939-z},
}

@article{wang2022respecting,
  title = {When and why {PINNs} fail to train: A neural tangent kernel perspective},
  author = {Wang, Sifan and Yu, Xinling and Perdikaris, Paris},
  journal = {Journal of Computational Physics},
  volume = {449},
  pages = {110768},
  year = {2022},
  publisher = {Elsevier},
  doi = {10.1016/j.jcp.2021.110768},
}
	
\end{document}